\documentclass[10pt]{iopart}
\usepackage{graphicx}
\eqnobysec

\newcommand{\beq}{\begin{equation}}
\newcommand{\beqa}{\begin{eqnarray}}
\newcommand{\eeq}{\end{equation}}
\newcommand{\eeqa}{\end{eqnarray}}

\renewcommand{\a}{\alpha}
\newcommand{\abs}[1]{\vert#1\vert}
\renewcommand{\bar}[1]{{\overline{#1}}}

\newcommand{\comport}[2]{\mathrel{\mathop{#1}\limits_{#2}^{\null}}}
\newcommand{\ket}[1]{\vert#1\rangle}
\newcommand{\braket}[2]{\langle#1\vert#2\rangle}
\newcommand{\braopket}[3]{\langle#1\vert#2\vert#3\rangle}
\newcommand{\cm}{{\rm cm}}
\renewcommand{\d}{{\rm d}}
\newcommand{\dif}[2]{{\frac{\d #1}{\d #2}}}
\newcommand{\eps}{\varepsilon}
\newcommand{\g}{\gamma}
\newcommand{\h}{\widehat}
\newcommand{\half}{\frac{1}{2}}
\newcommand{\ii}{{\rm i}}
\renewcommand{\max}{{\rm max}}

\newcommand{\mean}[1]{\langle#1\rangle}
\renewcommand{\o}{\omega}
\newcommand{\p}{\psi}
\newcommand{\q}{{q_\star}}
\newcommand{\tin}{{\widetilde{\tau}}}

\newcommand{\Ai}{{\rm Ai}}
\newcommand{\B}{{(\rm B)}}
\newcommand{\E}{{\cal E}}
\newcommand{\F}{{(\rm F)}}

\renewcommand{\Im}{\mathop{\rm Im}\nolimits}
\renewcommand{\Re}{\mathop{\rm Re}\nolimits}

\begin{document}

\title{Interacting quantum walkers: Two-body bosonic and fermionic bound states}

\author{P L Krapivsky$^{1,2}$, J M Luck$^2$, and K Mallick$^2$}

\address{$^1$ Department of Physics, Boston University, Boston, MA 02215, USA}

\address{$^2$ Institut de Physique Th\'eorique, Universit\'e Paris-Saclay, CEA and CNRS,
91191 Gif-sur-Yvette, France}

\begin{abstract}
We investigate the dynamics of bound states of two interacting particles,
either bosons or fermions,
performing a continuous-time quantum walk on a one-dimensional lattice.
We consider the situation where the distance between both particles
has a hard bound,
and the richer situation where the particles are bound by a smooth confining potential.
The main emphasis is on the velocity characterizing the ballistic spreading
of these bound states,
and on the structure of the asymptotic distribution profile
of their center-of-mass coordinate.
The latter profile generically exhibits many internal fronts.
\end{abstract}

\eads{\mailto{pkrapivsky@gmail.com},\mailto{jean-marc.luck@cea.fr},\mailto{kirone.mallick@cea.fr}}

\maketitle

\section{Introduction}

Quantum walks have witnessed an upsurge of interest in parallel
with the developments of quantum algorithms and quantum information
(see~\cite{kem,amb,sev} for reviews).
Two different types of quantum-mechanical analogues of classical random walks
have been investigated.
Discrete-time quantum walkers~\cite{adz,nk1,nk2,mbs} possess,
besides their spatial position, a finite-dimensional internal degree of freedom,
referred to as a quantum coin.
Both spatial and internal degrees of freedom jointly undergo a unitary dynamics.
Continuous-time quantum walkers~\cite{fg,tl,bbt} have no internal degree of freedom.
Their dynamics is governed by some hopping operator on the underlying structure.
Despite these differences,
continuous-time quantum walks can be viewed as a limit of discrete-time quantum
walks~\cite{fs}.
Both types of quantum walks exhibit many similar properties.
Their main characteristic feature is a fast ballistic spreading.
The typical distance traveled by a quantum walker grows linearly in time,
as opposed to the diffusive spreading of a classical random walker.

If two or more quantum walkers are simultaneously present,
the combined effects of interactions, quantum statistics and entanglement
give rise to novel features which have no classical counterpart.
The Anderson localization of two interacting quantum particles in a random potential
has attracted much attention~\cite{ds,yi,kf}.
More recently, dynamical features of the quantum walks performed
by two or more entangled or interacting particles
have been the subject of numerous theoretical studies
(see e.g.~\cite{ops,gfz,bw,mtm,rss,sbk,aam,lvh,b3,cb,qkg}).
Several experimental groups have also studied the quantum walk
of entangled pairs of magnons~\cite{fse}
and of photons in various integrated photonics devices~\cite{plm,ssv,cor,preiss}.

Here we investigate the continuous-time quantum walk
of bound states of two identical bosonic or fermionic particles.
The main emphasis is on the distribution profile
in the center-of-mass coordinate of the bound states,
including the dependence of the ballistic spreading velocity on model parameters,
and the generic presence of many internal ballistic fronts.
The setup of this paper is as follows.
In section~\ref{sone} we revisit in a pedagogical way
the continuous-time quantum walk of a single particle.
We analyze the distribution of the position of the particle,
emphasizing its dependence on the initial quantum state.
We also show that allowing the particle to hop to second neighbors
may yield a novel effect, namely the appearance of internal ballistic fronts in
the position distribution, besides the usual extremal fronts.
We then address similar questions in the more challenging situation
of the quantum walk performed by bound states of two identical particles.
The same formalism allows one to deal with bosonic and fermionic bound states,
as they are respectively described by even and odd functions
of the relative coordinate between both particles.
We consider bound states generated either by imposing
a hard bound on the distance between both particles (section~\ref{shard})
or by a smooth confining potential (section~\ref{ssoft}).
In both situations the emphasis is on the distribution of the center-of-mass coordinate
and on the presence of internal ballistic fronts.
Section~\ref{discussion} contains a discussion of our findings.

\section{Quantum walk of a single particle}
\label{sone}

This section serves as a self-contained pedagogical introduction
to the main concepts emphasized in the rest of the paper.
We consider the continuous-time quantum walk
performed by a single particle on a one-dimensional lattice.
Most of the features underlined in this section have been studied,
or at least mentioned, in several earlier works
in the context of discrete-time quantum walks~\cite{sbk,bca,tfm,iks,mkk,fb1,tm2,sbj}.
One of the purposes of this section is to demonstrate that the analysis
of the continuous-time quantum walk problem is much simpler,
and can therefore be worked out in a more systematic fashion.

More specifically,
we first revisit the simple quantum walk where the particle hops to nearest
neighbors only,
focusing our attention onto the asymptotic distribution
of the walker, including its dependence on the initial state.
We then turn to a generalized quantum walk,
where the particle hops to further neighbors as well.
We show that allowing hopping to second neighbors gives rise
to internal fronts in the distribution profile.
Hopping to second neighbors is known to
have drastic physical consequences in many other situations.
One celebrated example is graphene (see~\cite{g1,g2}),
where hopping to second neighbors breaks the chiral symmetry between both sublattices.

\subsection{Simple quantum walk}
\label{smin}

The framework of the simple quantum walk
is the usual one of the tight-binding approximation,
where the particle hops to nearest neighbors only.
Throughout this paper, we use dimensionless units.
The wavefunction $\p_n(t)=\braket{n}{\p(t)}$ of the particle at site~$n$ at
time~$t$ obeys
\beq
\ii\,\dif{\p_n(t)}{t}=\p_{n+1}(t)+\p_{n-1}(t).
\label{dif}
\eeq
The dispersion law between wavevector (momentum) $q$ and frequency (energy)
$\o$
and the corresponding group velocity $v$ therefore read
\beq
\o(q)=2\cos q,\qquad v(q)=\o'(q)=-2\sin q,
\label{disp}
\eeq
where the prime denotes a derivative.

\subsection*{Initial state localized at the origin}

Suppose that at time $t=0$ the particle
is localized on a single site, taken as the origin of the lattice:
$\p_n(0)=\delta_{n,0}$.
In Fourier space, $\h\p(q,0)=1$, hence $\h\p(q,t)=\e^{-\ii\o(q)t}$, and so
\beq
\p_n(t)=\int_0^{2\pi}\frac{\d q}{2\pi}\,\e^{\ii(nq-2t\cos
q)}=\ii^{-n}\,J_n(2t),
\label{pint}
\eeq
where the $J_n$ are the Bessel functions.

\begin{figure}[!ht]
\begin{center}
\includegraphics[angle=-90,width=.5\linewidth]{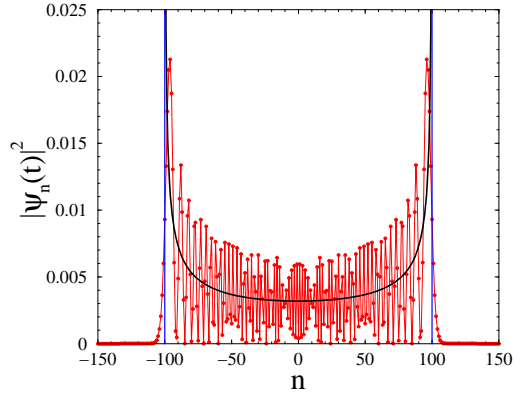}
\caption{\small
Probability profile $\abs{\p_n(t)}^2$ at time $t=50$
for the simple quantum walk with a particle initially located at the origin.
Vertical blue lines: theoretical positions of the ballistic fronts.
Black curve: asymptotic distribution~(\ref{f1}) set to scale.}
\label{min}
\end{center}
\end{figure}

Figure~\ref{min} shows the probabilities $\abs{\p_n(t)}^2=(J_n(2t))^2$
against position $n$ at time $t=50$.
These probabilities exhibit various regimes of behavior at large $n$ and $t$.
They take appreciable values in the allowed region
which spreads ballistically with the maximal velocity
\beq
V=\comport{\max}{q}\abs{\o'(q)}=2.
\eeq
Furthermore, they exhibit sharp fronts near $n=\pm2t$, and they decay exponentially
in the forbidden region beyond these fronts.
These asymptotic results are classics
of the theory of Bessel functions~\cite{watson,bateman},
whose physical meaning has been underlined in~\cite{tl}.
These results can be readily obtained the saddle-point method,
which will be used later in other situations.

\begin{itemize}

\item
{\it Allowed region} ($\abs{n}<2t$).
In this region, the oscillatory behavior of the wave\-function at long times
can be derived as follows.
Setting
\beq
n=-2t\sin\q\qquad\left(\abs{\q}<\pi/2\right),
\label{nsp}
\eeq
the integral in~(\ref{pint})
is dominated by two equivalent saddle points at $q=\q$ and $q=\pi-\q$ (modulo $2\pi$).
We thus recover the well-known asymptotic form of Bessel functions:
\beq
J_n(2t)=\ii^n\p_n(t)
\approx\frac{\cos(2t\cos\q-n(\pi/2+\q)+\pi/4)}{\sqrt{\pi t\cos\q}}.
\eeq
By averaging out the oscillations in the probabilities $\abs{\p_n(t)}^2$, one
arrives~\cite{tl} at a smooth distribution $f(v)$ for the ratio
$v=n/t$ in the long-time limit, i.e.,
\beq
f(v)=\frac{1}{\pi\sqrt{4-v^2}}\qquad(\abs{v}<2).
\label{f1}
\eeq
The above result can be alternatively obtained
by folding the uniform distribution on the Brillouin zone $[-\pi,\pi]$
by the dispersion curve of the group velocity: $v=-2\sin q$.

\item
{\it Transition region} ($\abs{n}\approx 2t$).
The distribution~(\ref{f1}) becomes singular as the endpoints of the allowed
region are approached ($n\to\pm2t$).
The vicinity of these endpoints corresponds to the transition region
in the theory of Bessel functions.
Setting $\abs{n}=2t+zt^{1/3}$, we have
\beq
J_n(2t)=\ii^n\p_n(t)\approx t^{-1/3}\Ai(z),
\label{jtrans}
\eeq
where $\Ai(z)$ is the Airy function.
The probabilities $\abs{\p_n(t)}^2$
therefore exhibit sharp ballistic fronts~\cite{tl},
whose height and width respectively scale as $t^{-2/3}$ and~$t^{1/3}$.

\item
{\it Forbidden region} ($\abs{n}>2t$).
In this region, the exponential fall-off of the wavefunction at long times
can be derived by evaluating again the integral in~(\ref{pint}) by the saddle-point method.
Setting $\abs{n}=2t\cosh\theta$ with $\theta>0$, we obtain
\beq
J_n(2t)=\ii^n\p_n(t)\approx\frac{\e^{-2t(\theta\cosh\theta-\sinh\theta)}}
{\sqrt{4\pi t\sinh\theta}}.
\eeq

\end{itemize}

\subsection*{Arbitrary initial state}

For an arbitrary initial state $\p_n(0)$, we have
\beqa
\p_n(t)&=&\int_0^{2\pi}\frac{\d q}{2\pi}\,\e^{\ii(nq-2t\cos q)}\h\p(q,0)
\nonumber\\
&=&\sum_m\ii^{m-n}\,\p_m(0) J_{n-m}(2t).
\label{gpint}
\eeqa

Whenever the initial state is reasonably localized,
in the sense that $\p_n(0)$ decays fast enough with distance $\abs{n}$,
the time-dependent wavefunction exhibits qualitatively the regimes of behavior
described above in the allowed region, ballistic fronts, and forbidden region.

Let us analyze more carefully the allowed region ($\abs{n}<2t$).
With the definition~(\ref{nsp}), the integral in~(\ref{gpint})
is now dominated by two inequivalent saddle points at $q=\q$ and $q=\pi-\q$
(modulo $2\pi$).
We thus arrive at
\beqa
\p_n(t)\approx\frac{1}{\sqrt{4\pi\cos\q}}
\Bigl(&&\h\p(\q,0)\,\e^{\ii(n\q-2t\cos\q+\pi/4)}
\nonumber\\
&&+\h\p(\pi-\q,0)\,\e^{\ii(n(\pi-\q)+2t\cos\q-\pi/4)}\Bigr).
\eeqa

Averaging out the oscillations in the above expression yields the following
formula for the locally coarse-grained probabilities:
\beq
\abs{\p_n(t)}^2\approx\frac{\abs{\h\p(\q,0)}^2+\abs{\h\p(\pi-\q,0)}^2}{4\pi t\cos\q}.
\label{pgen}
\eeq
The limit distribution $f(v)$ of the ratio $v=n/t$ is obtained by folding
the distribution on the Brillouin zone with density $\abs{\h\p(q)}^2/(2\pi)$
by the dispersion curve $v=-2\sin q$.
This line of thought has been used in~\cite{fb1,tm2},
and more thoroughly in~\cite{sbj}, for discrete-time walks.

For definiteness, let us consider the case where the initial wavefunction is
spread over three consecutive sites, i.e.,
\beq
\p_n(0)=a\delta_{n,1}+b\delta_{n,0}+c\delta_{n,-1}
\qquad(\abs{a}^2+\abs{b}^2+\abs{c}^2=1).
\label{three}
\eeq
We have
\beq
\h\p(q,0)=a\,\e^{-\ii q}+b+c\,\e^{\ii q}
\eeq
and
\beq
\p_n(t)=\ii^{-n}\bigl(bJ_n(2t)+\ii(aJ_{n-1}(2t)-cJ_{n+1}(2t))\bigr).
\eeq
The asymptotic result~(\ref{pgen}) reads explicitly
\beq
\abs{\p_n(t)}^2\approx\frac{1+4A\cos^2\q-2B\sin\q}{2\pi t\cos\q},
\eeq
with the definition~(\ref{nsp}), and where\footnote{The bar denotes complex conjugation.}
\beq
A=\Re(a\bar{c}),\qquad B=\Im(b(\bar{a}-\bar{c})).
\eeq
The limit distribution $f(v)$ of the ratio $v=n/t$ reads therefore
\beq
f(v)=\frac{1+A(2-v^2)+Bv}{\pi\sqrt{4-v^2}}\qquad(\abs{v}<2).
\label{fgen}
\eeq
Analogous expressions have been derived in~\cite{nk1,nk2,fb1,tm2,sbj}
for various models of discrete-time quantum walks.
The distribution~(\ref{fgen}) generically has
inverse-square-root singularities at the endpoints of the allowed region
($v=\pm2$), corresponding to the ballistic fronts.
The initial state only affects the numerator (see~(\ref{f1})).
This lack of universality, namely the ever-lasting memory of the initial state,
is a genuine quantum feature
(it is absent for the classical walker with sufficiently localized initial state).

The first moments of the distribution~(\ref{fgen}),
\beq
\mean{v}=2B,\qquad\mean{v^2}=2(1-A),
\eeq
match the asymptotic growth at long times of the exact expressions
\beqa
\mean{n^k(t)}
&=&\braopket{\p(t)}{n^k}{\p(t)}
\nonumber\\
&=&\int_0^{2\pi}\frac{\d q}{2\pi}\,\bar{\h\p(q,0)}\,\e^{2\ii t\cos q}
\left(\ii\frac{\d}{\d q}\right)^k\e^{-2\ii t\cos q}\h\p(q,0)
\label{moms}
\eeqa
of the position moments~\cite{tl}, namely
\beqa
\mean{n(t)}&=&\abs{a}^2-\abs{c}^2+2Bt,
\nonumber\\
\mean{n^2(t)}&=&\abs{a}^2+\abs{c}^2+2\Im(b(\bar{a}+\bar{c}))t+2(1-A)t^2.
\eeqa

\begin{figure}[!ht]
\begin{center}
\includegraphics[angle=-90,width=.47\linewidth]{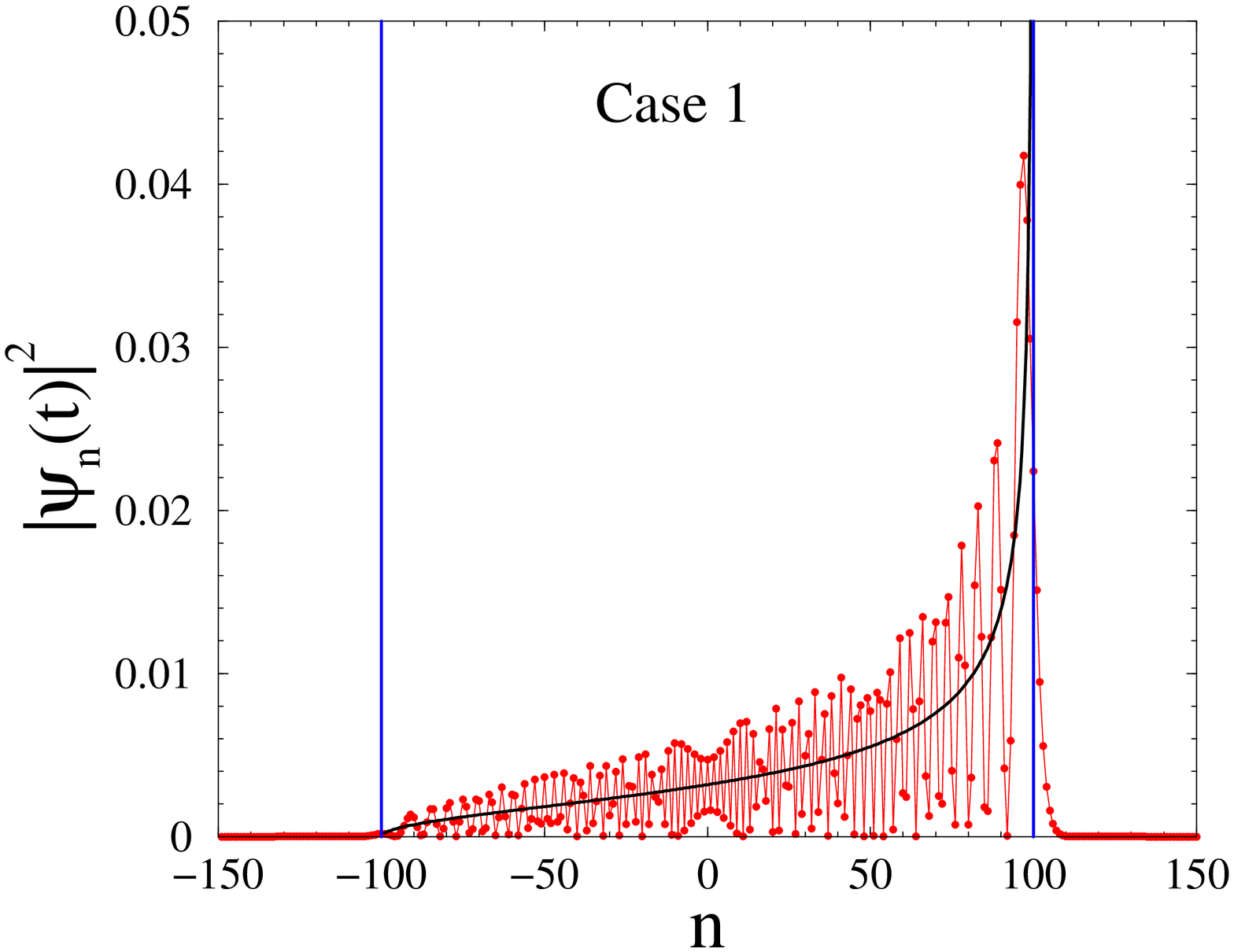}
\hskip 6pt
\includegraphics[angle=-90,width=.47\linewidth]{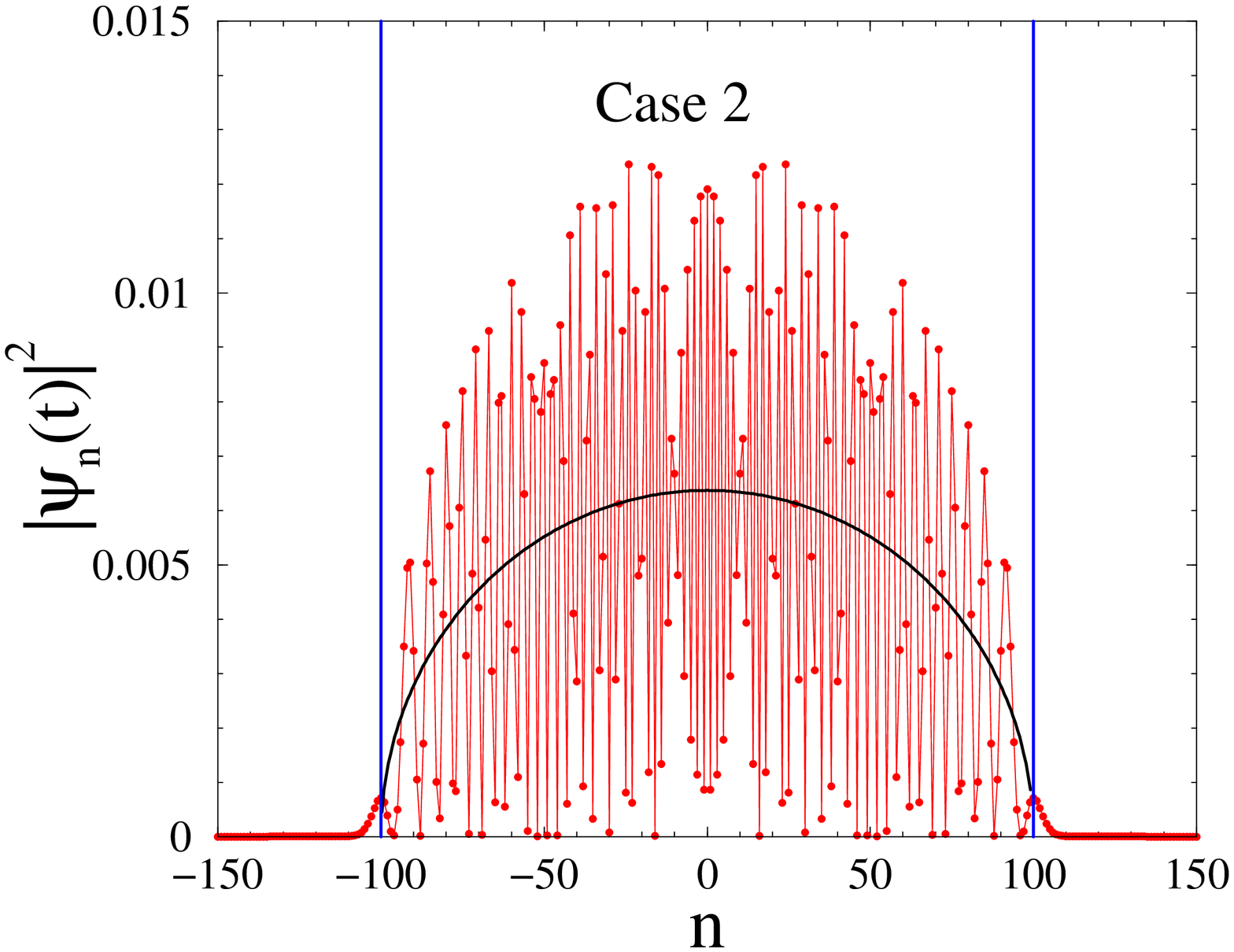}
\caption{\small
Probability profiles $\abs{\p_n(t)}^2$ at time $t=50$
for the simple quantum walk with the two exceptional
initial states described in the text.
Vertical blue lines: theoretical positions of the ballistic fronts.
Black curves: asymptotic distributions~(\ref{fone}),~(\ref{ftwo}) set to
scale.}
\label{pmin}
\end{center}
\end{figure}

The inverse-square-root singularities at the endpoints of the allowed region ($v=\pm2$)
are generic but not entirely universal.
There are indeed special initial states such that either one or both endpoint singularities
are absent from the limit distribution~(\ref{fgen}).
Figure~\ref{pmin} shows the probability profiles $\abs{\p_n(t)}^2$
at time $t=50$ in the following two special cases of initial states.

\begin{itemize}

\item[1.]
For $a=1/\sqrt2$, $b=\ii/\sqrt2$ and $c=0$, hence $A=0$ and $B=1/2$, we have
\beq
\p_n(t)=\frac{\ii^{1-n}}{\sqrt{2}}\bigl(J_{n-1}(2t)+J_n(2t)\bigr),
\eeq
\beq
f(v)=\frac{1}{2\pi}\sqrt{\frac{2+v}{2-v}}\qquad(\abs{v}<2),
\label{fone}
\eeq
and so the left front is absent (see figure~\ref{pmin}, left).

\item[2.]
For $a=c=1/\sqrt2$ and $b=0$, hence $A=1/2$ and $B=0$, we have
\beq
\p_n(t)=\frac{\ii^{1-n}}{\sqrt{2}}\bigl(J_{n-1}(2t)-J_{n+1}(2t)\bigr),
\eeq
\beq
f(v)=\frac{\sqrt{4-v^2}}{2\pi}\qquad(\abs{v}<2),
\label{ftwo}
\eeq
and so both fronts are absent (see figure~\ref{pmin}, right).

\end{itemize}

This kind of non-generic behavior of the distribution profile
has also been observed for discrete-time quantum walks,
both with the usual two-dimensional internal state~\cite{tfm,sbk}
and with a more exotic three-dimensional quantum coin
which may lead to a localization phenomenon,
in the sense that part of the probability stays forever
at a finite distance from the particle's starting point~\cite{iks,sbj}.
Reference~\cite{sbj} also contains analytical predictions for $f(v)$,
somehow similar to~(\ref{fone}) and~(\ref{ftwo}),
in many special situations.

\subsection{Generalized quantum walk}
\label{sext}

Novel phenomena occur in generalized quantum walks,
where the particle may hop to further neighbors.
Let us consider the minimal generalization where hops are limited
to first (nearest) and second (next-nearest) neighbors,
with respective amplitudes 1 and $g$.
The wavefunction of the particle at site~$n$ at time~$t$ now obeys
\beq
\ii\,\dif{\p_n(t)}{t}=\p_{n+1}(t)+\p_{n-1}(t)+g\left(\p_{n+2}(t)+\p_{n-2}(t)\right).
\label{edif}
\eeq
We have therefore
\beqa
\o(q)&=&2(\cos q+g\cos 2q),
\nonumber\\
\o'(q)&=&v(q)=-2(\sin q+2g\sin 2q),
\nonumber\\
\o''(q)&=&-2(\cos q+4g\cos 2q).
\label{edisp}
\eeqa
The above dispersion law $\o(q)$ is invariant under the transformation
$g\to-g$, $q\to q+\pi$, $\o\to-\o$.
We may therefore restrict the analysis to the domain $g\ge0$,
without any loss of generality.

Suppose the particle is launched from the origin: $\p_n(0)=\delta_{n,0}$.
Thus $\h\p(q,0)=1$, and so
\beq
\p_n(t)=\int_0^{2\pi}\frac{\d q}{2\pi}\,\e^{\ii(nq-\o(q)t)}.
\label{epint}
\eeq

Some observables have a smooth dependence on the amplitude $g$.
This is especially the case for the position moments (see~(\ref{moms})), which read
\beqa
\mean{n^2(t)}&=&2(1+4g^2)t^2,
\nonumber\\
\mean{n^4(t)}&=&2(1+16g^2)t^2+6(1+16g^2+16g^4)t^4,
\eeqa
and so on (odd moments vanish identically by symmetry).

The presence of hopping to second neighbors however introduces a novel
qualitative feature.
Let us for the time being adopt a general standpoint,
and consider the wavefunction~(\ref{epint}) in the regime of long times.
Evaluating the integral by the saddle-point method,
as we did in section~\ref{smin}, we obtain
\beq
\p_n(t)\approx\sum_q\frac{\e^{\ii(nq-\o(q)t)}}{\sqrt{2\pi\ii\o''(q)t}},
\eeq
where the sum runs over the solutions $q$ of the saddle-point equation
\beq
\frac{n}{t}=v=\o'(q).
\eeq
By averaging out the oscillations in the probabilities $\abs{\p_n(t)}^2$,
we again predict that $v$ has a smooth distribution
\beq
f(v)=\frac{1}{2\pi}\sum_q\frac{1}{\abs{\o''(q)}}
\label{ef}
\eeq
in the long-time limit.
In full generality, the above distribution is again obtained
by folding the uniform distribution on the Brillouin zone
by the dispersion map of the group velocity $v=\o'(q)$,
in the sense that $\d q/(2\pi)=f(v)\d v$ holds formally.

The distribution~(\ref{ef}) has generically inverse-square-root singularities
at all extremal values of $v$,
in correspondence with wavevectors $q$ so that $\d v/\d q=\o''(q)$ vanishes.
It is always singular at the endpoints of the allowed region ($v=\pm V$),
as the maximal velocity $V$ is necessarily an extremal value.
The distribution~(\ref{ef}) may however also have internal singularities
within the allowed region,
which were absent in the simple quantum walk considered in section~\ref{smin}.

Internal ballistic fronts of that kind
have been observed in two generalizations of the usual
discrete-time quantum walk~\cite{bca,mkk}.
Reference~\cite{bca} investigates a quantum walk
subjected to $M$ independent quantum coins acting cyclically.
If $M$ is large,
the dynamics exhibits a crossover between classical random walk at short times ($t\ll M$)
and quantum walk at long times ($t\gg M$).
In the latter regime the distribution of the quantum particle
exhibits an array of equally spaced ballistic peaks,
whose number grows as~$M/2$,
the distance between any two consecutive peaks being $\Delta v=\sqrt{2}/M$.
The model studied in~\cite{mkk} is closer to ours in its spirit.
It consists of a discrete-time quantum walk where hops up to distance $j$ are allowed,
whose dynamics is rigidly
dictated by $(2j+1)$-dimensional Wigner rotation matrices.
Here too, the distribution profile exhibits $2j+1$ equally spaced ballistic peaks.

In order to pursue the analysis,
let us specialize to the continuous-time quantum walk
with hops to first and second neighbors only (see~(\ref{edif}),~(\ref{edisp})).
This minimal example is already too complex to allow one to turn~(\ref{ef})
to an explicit expression.
The location of the ballistic peaks can nevertheless be predicted as follows.
The second derivative $\o''(q)$ vanishes for
\beq
\cos q_\pm=\frac{-1\pm(1+128g^2)^{1/2}}{16g}.
\eeq
The corresponding values $V_\pm$ of $v$ are given by
\beq
V_\pm^2=\frac{-1+320g^2+2048g^4\pm(1+128g^2)^{3/2}}{128g^2}.
\eeq
The allowed region always spreads ballistically with the maximal velocity $V_+$.
The smaller velocity $V_-$ may also play a role, depending on the strength of $g$:

\begin{itemize}

\item
If the amplitude $g$ is small enough ($g<g_c=1/4$),
the situation is qualitatively similar to that of the simple quantum walk,
studied in section~\ref{smin}.
We have indeed $\cos q_-<-1$ and $0<\cos q_+<1$, so that only $q_+$ matters,
and so $f(v)$ is only singular at the endpoints $\pm V_+$ of the allowed region.

\item
If the amplitude $g$ is large enough ($g>g_c=1/4$), both $q_+$ and $q_-$ matter.
As a consequence, the distribution $f(v)$
has two singularities at the endpoints $\pm V_+$
and two internal singularities at the smaller values $\pm V_-$.
The wavefunction exhibits four ballistic fronts:
two extremal ones, propagating at the maximal velocity ($\pm V_+$),
and two internal ones, propagating at a smaller velocity ($\pm V_-$).

\item
In the borderline case ($g=g_c=1/4$),
we have $\cos q_-=-1$, i.e., $q_-=\pi$ and $V_-=0$.
The dispersion curve exhibits an unusual quartic behavior near $q=\pi$:
\beq
\o(q)\approx-\frac{3}{2}+\frac{\eps^4}{4}\qquad(\eps=q-\pi\to0).
\eeq
This anomalous dispersion right at $g=g_c$ has two consequences.
First, the distribution $f(v)$ has a singularity at $v=0$, of the form
\beq
f(v)\approx\frac{1}{6\pi\abs{v}^{2/3}}.
\label{fvc}
\eeq
This central singularity with exponent $-2/3$ is stronger than the generic ones,
whose exponent is $-1/2$.
Second, the wavefunction at the origin exhibits an unusually slow fall-off:
\beq
\p_0(t)
\approx\frac{\sqrt{2}\,\Gamma(5/4)\,\e^{\ii(3t/2-\pi/8)}}{\pi t^{1/4}}.
\eeq
This $t^{-1/4}$ decay is slower than the generic $t^{-1/2}$ decay
exhibited e.g.~by the Bessel function $J_0(2t)$ (see~(\ref{pint})).

\end{itemize}

\begin{figure}[!ht]
\begin{center}
\includegraphics[angle=-90,width=.5\linewidth]{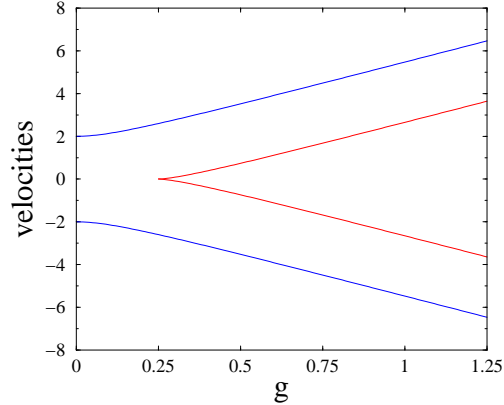}
\caption{\small
Front velocities of the quantum walk with hopping to first and second neighbors
(see~(\ref{edif})), against the amplitude~$g$.
External (blue) curves: velocities $\pm V_+$ of the extremal fronts
(endpoints of allowed region).
Internal (red) curves: velocities $\pm V_-$ of the internal fronts
for $g>g_c=1/4$.}
\label{vext}
\end{center}
\end{figure}

Figure~\ref{vext} shows the dependence of the front velocities against $g$
for $g\ge0$.
The larger velocities $\pm V_+$ of the extremal fronts
(blue curves) describe the endpoints of the allowed region for all $g$.
The smaller velocities $\pm V_-$ of the internal fronts
(red curves) exist for $g>g_c=1/4$ only.
At $g=g_c$ we have $V_+=3\sqrt{3}/2$,
while $V_-$ takes off as $(32\sqrt{6}/9)(g-g_c)^{3/2}$.
At large $g$, both velocities grow linearly in $g$ with the same slope,
as $V_\pm\approx4g\pm\sqrt{2}$.
This non-trivial pattern of front velocities is richer than the periodic arrays
of equally spaced peaks observed in~\cite{bca,mkk}.

\begin{figure}[!ht]
\begin{center}
\includegraphics[angle=-90,width=.47\linewidth]{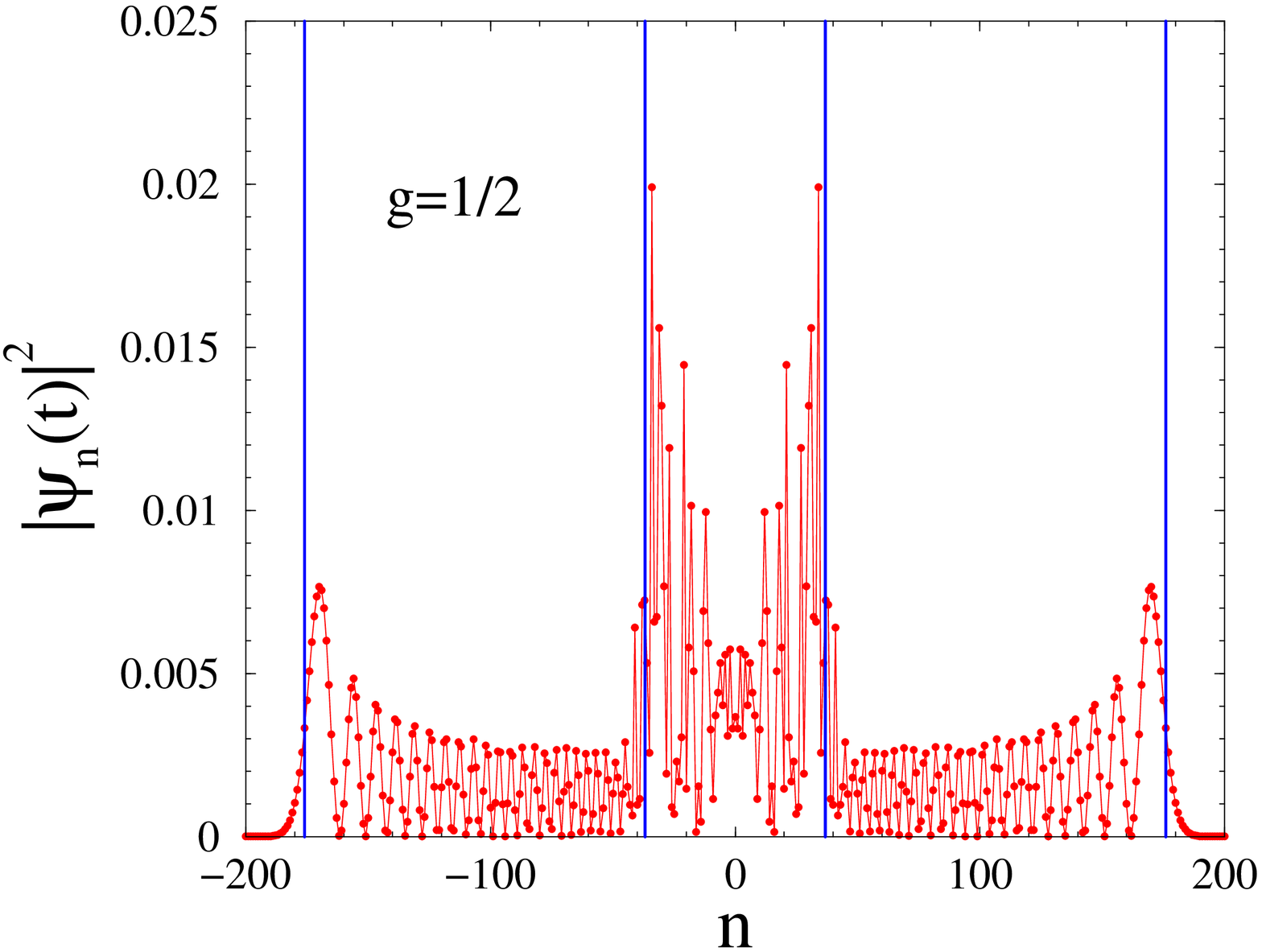}
\hskip 6pt
\includegraphics[angle=-90,width=.47\linewidth]{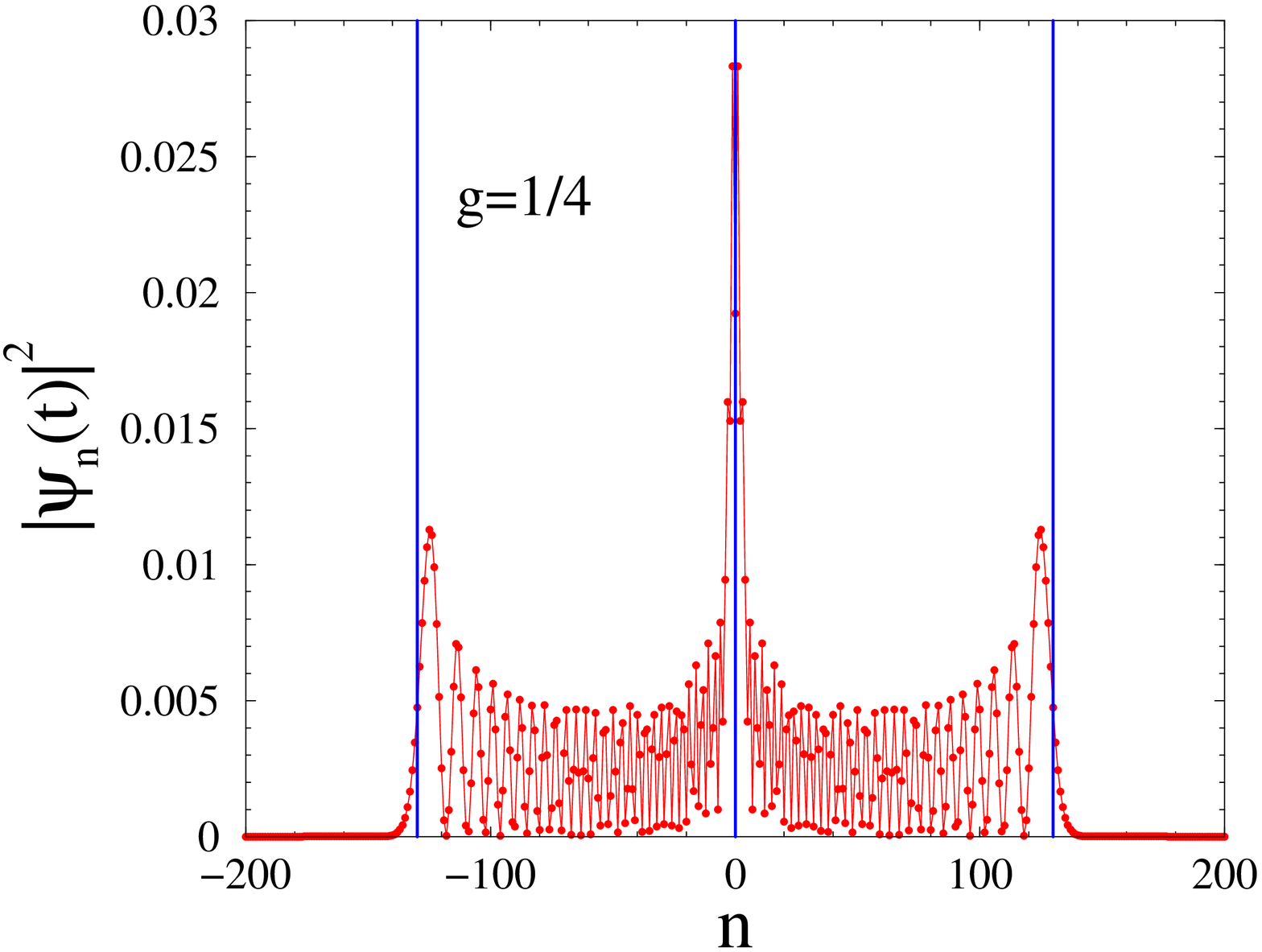}
\caption{\small
Probabilities $\abs{\p_n(t)}^2$ for the quantum walk~(\ref{edif}) at time $t=50$.
Vertical blue lines: theoretical positions of the ballistic fronts.
Left ($g=1/2$): four fronts at $\pm V_+$ and $\pm V_-$.
Right ($g=g_c=1/4$): two fronts at $\pm V_+$ and a central one at the origin,
corresponding to the anomalous power-law singularity~(\ref{fvc}).}
\label{pext}
\end{center}
\end{figure}

Figure~\ref{pext} shows the probabilities $\abs{\p_n(t)}^2$
at time $t=50$ for a particle launched at the origin and two values of $g$.
For $g=1/2$ (left), the probability profile exhibits four fronts.
For $g=g_c=1/4$ (right), the probability profile exhibits three fronts.
The central one at the origin corresponds to the anomalous singularity~(\ref{fvc}).

\section{Two-body bound states: hard bound on distance}
\label{shard}

We now turn to our main subject: the quantum walk
performed by a bound state of two identical particles
propagating coherently along a one-dimensional lattice.
The main emphasis will be on the asymptotic distribution
of the center-of-mass coordinate of the bound state,
on the velocity characterizing its ballistic spreading
and on the structure of the distribution profile,
which generically exhibits many internal fronts.
To the best of our knowledge,
these matters have only been addressed so far in two papers~\cite{aam,qkg}.
In order not to interrupt the lengthy developments of sections~\ref{shard} and~\ref{ssoft},
we postpone the discussion of those earlier works to section~\ref{discussion}.

Here again, considering continuous-time walks will
allow for a more thorough and systematic investigation of the problem.
Bound states obtained by imposing a hard bound $\ell$
on the distance between both particles are dealt with in this section,
whereas those generated by a smooth confining potential
will be considered in section~\ref{ssoft}.
The same formalism will allow one to deal with bosonic and fermionic bound states,
as they are respectively described by even and odd functions
of the relative coordinate~$m$ between both particles.

\subsection{Generalities}

\begin{figure}[!ht]
\begin{center}
\includegraphics[angle=-90,width=.5\linewidth]{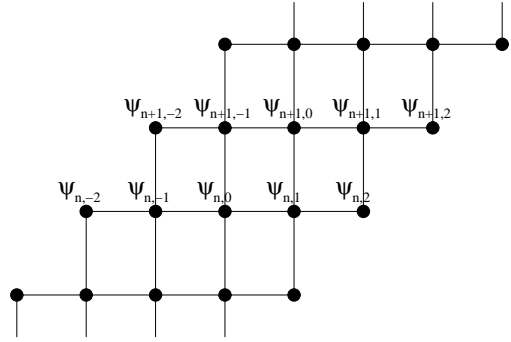}
\caption{\small
Plot of the positions of the particles in the $(n_1,n_2)$ plane for $\ell=2$.
Full symbols denote the allowed configurations of the particles.
Links between symbols show the allowed hops of any of the particles to a
neighboring site.}
\label{stair}
\end{center}
\end{figure}

We denote by $n_1=n+m$ and $n_2=n$ the abscissas of two identical particles on the lattice,
where $m=n_1-n_2$ is the relative coordinate.
We impose a hard bound $\ell$ on the distance $\abs{m}$ between both particles,
and so $m$ is restricted to the $2\ell+1$ values $m=-\ell,\dots,\ell$.
Figure~\ref{stair} shows the positions of the particles
in the $(n_1,n_2)$ plane for $\ell=2$.
Full symbols denote the allowed configurations of the particles.
Links between symbols show the allowed hops of any of the particles to a neighboring site.

The time-dependent wavefunction
\beq
\p_{n,m}(t)=\braket{(n_1,n_2)}{\p(t)}=\braket{(n+m,n)}{\p(t)}
\label{2psi}
\eeq
obeys the equation
\beqa
\ii\,\dif{\p_{n,m}(t)}{t}&=&\p_{n,m-1}(t)+\p_{n+1,m-1}(t)
\nonumber\\
&+&\p_{n-1,m+1}(t)+\p_{n,m+1}(t),
\label{2dif}
\eeqa
with boundary conditions $\p_{n,\pm(\ell+1)}=0$.

\subsection{Bosonic and fermionic spectra}

A basis of plane-wave solutions to~(\ref{2dif}) reads
\beq
\p_{n,m}(t)=\e^{\ii(n_\cm q+mp-\o(p,q)t)},
\label{pw}
\eeq
where the momenta $p$ and $q$ are respectively conjugate to
the relative coordinate $m=n_1-n_2$ and to the center-of-mass coordinate
\beq
n_\cm=\frac{n_1+n_2}{2}=n+\frac{m}{2}.
\label{ncm}
\eeq
The resulting dispersion relation has a product form~\cite{qkg}:
\beq
\o(p,q)=4\cos p\cos\frac{q}{2}.
\label{wcont}
\eeq

\begin{itemize}

\item
{\it Bosonic states} are described by even functions
under the exchange of $n_1$ and $n_2$; they are
obtained by adding the plane waves~(\ref{pw}) for $p$ and $-p$:
\beq
\p_{n,m}^\B(t)=\e^{\ii(n_\cm q-\o(p,q)t)}\cos mp.
\label{pb}
\eeq
The relative momentum $p$ is quantized by the condition $\cos(\ell+1)p=0$.
It therefore takes the $\ell+1$ values
\beq
p_k^\B=\frac{(k+\half)\pi}{\ell+1}\qquad(k=0,\dots,\ell),
\eeq
and so
\beq
\o_k^\B(q)=4\cos\frac{(k+\half)\pi}{\ell+1}\,\cos\frac{q}{2}.
\label{wb}
\eeq
The bosonic dispersion curve thus consists of $\ell+1$ branches, with group velocities
\beq
v_k^\B(q)=-2\cos\frac{(k+\half)\pi}{\ell+1}\,\sin\frac{q}{2}.
\label{vb}
\eeq

\item
{\it Fermionic states} are described by odd functions of $m$,
obtained by subtracting the plane waves~(\ref{pw}) for $p$ and $-p$:
\beq
\p_{n,m}^\F(t)=\e^{\ii(n_\cm q-\o(p,q)t)}\sin mp.
\label{pf}
\eeq
As fermionic particles cannot cross each other in one dimension,
the range of the fermionic wavefunctions~(\ref{pf})
can be restricted to the sector $n_1>n_2$, i.e., $m=1,\dots,\ell$.
The relative momentum $p$ is quantized by the condition $\sin(\ell+1)p=0$.
It therefore takes the $\ell$ values
\beq
p_k^\F=\frac{k\pi}{\ell+1}\qquad(k=1,\dots,\ell),
\eeq
and so
\beq
\o_k^\F(q)=4\cos\frac{k\pi}{\ell+1}\,\cos\frac{q}{2}.
\label{wf}
\eeq
The fermionic dispersion curve thus consists of $\ell$ branches, with group velocities
\beq
v_k^\F(q)=-2\cos\frac{k\pi}{\ell+1}\,\sin\frac{q}{2}.
\label{vf}
\eeq

\end{itemize}

\begin{figure}[!ht]
\begin{center}
\includegraphics[angle=-90,width=.5\linewidth]{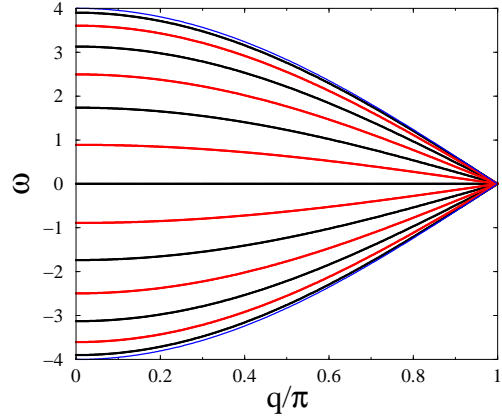}
\caption{\small
Plot of the bosonic and fermionic spectra
for a maximal distance $\ell=6$, against $q/\pi$.
Black: bosonic frequencies $\o_k^\B(q)$.
Red: fermionic frequencies $\o_k^\F(q)$.
Blue: band boundaries ($\o=\pm4\cos(q/2)$) (see~(\ref{wcont})).}
\label{whard}
\end{center}
\end{figure}

Figure~\ref{whard} shows the bosonic and fermionic spectra
for a maximal distance $\ell=6$.
The 7 bosonic frequencies $\o_k^\B(q)$ (black)
and the 6 fermionic frequencies $\o_k^\F(q)$ (red) are plotted against $q/\pi$.

\subsection{Ballistic spreading}

Let us now investigate the asymptotic behavior of the wavefunction $\p_{n,m}(t)$,
starting from an arbitrary initial state, where both particles are located in
the vicinity of the origin.
The detailed analysis of the one-body problem performed in section~\ref{sone}
allows one to draw the following picture.

The various components $\p_{n,m}(t)$ of the wavefunction spread ballistically,
i.e., their extension in the center-of-mass coordinate $n_\cm\approx n$
grows asymptotically linearly in time $t$ and symmetrically with respect to the origin.
The $\p_{n,m}(t)$ generically exhibit sharp fronts near $n=V_kt$,
where the front velocities $V_k$ are the stationary values of the group velocity,
corresponding to the boundaries of the Brillouin zone.
Setting $q=\pm\pi$ in~(\ref{vb}),~(\ref{vf}), we get
\beqa
V_k^\B&=&2\cos\frac{(k+\half)\pi}{\ell+1}\qquad(k=0,\dots,\ell),
\nonumber\\
V_k^\F&=&2\cos\frac{k\pi}{\ell+1}\qquad(k=1,\dots,\ell).
\label{vk}
\eeqa

In particular, all the components of the wavefunction
take appreciable values only in the allowed zone defined by $\abs{n}<Vt$,
where the maximal velocities
\beq
V^\B=2\cos\frac{\pi}{2(\ell+1)},\qquad
V^\F=2\cos\frac{\pi}{\ell+1},
\label{vbfres}
\eeq
respectively correspond to setting $k=0$ and $k=1$ in~(\ref{vk}).
Both maximal velocities approach the free value $V=2$
in the limit of a large confining size ($\ell\gg1$),
with two different correction amplitudes, i.e.,
\beq
V^\B\approx2-\frac{C^\B}{\ell^2},\qquad
V^\F\approx2-\frac{C^\F}{\ell^2},
\label{vhard}
\eeq
\beq
C^\B=\frac{\pi^2}{4},\qquad
C^\F=\pi^2.
\label{chard}
\eeq

For a generic initial state localized in the vicinity of the origin,
the various components $\p_{n,m}(t)$ of the wavefunction will exhibit,
besides two extremal ballistic fronts at $n=\pm V^\B t$ or $n=\pm V^\F t$,
$\ell-1$ internal fronts in the bosonic case (for $\ell\ge2$)
and $\ell-2$ internal fronts in the fermionic case (for $\ell\ge3$).

\subsection{Continuum limit and corrections}

Our bound-state problem owes its non-triviality
and its richness to the fact that particles live on a lattice.
In the continuum limit,
the dynamics of the center-of-mass coordinate and of the relative coordinate
are exactly decoupled, as a consequence of Galilean invariance.
As soon as the line is discretized into a lattice,
this Galilean invariance is broken.
This phenomenon has already been underlined
for discrete-time interacting quantum walkers~\cite{aam}.
Its consequences in the context of diluted Fermi gases and of the BCS--BEC crossover
have also been discussed recently~\cite{pc}.

In the present situation, too,
the dynamics of the center-of-mass coordinate~$n_\cm$ and of the relative coordinate $m$
are expected to decouple as the continuum limit is approached,
where Galilean invariance should be restored.
It is worth investigating in a quantitative way how this decoupling takes place,
starting from the lattice dispersion curves~(\ref{wb}) and~(\ref{wf}).
For that purpose, we introduce the lattice spacing~$a$,
and define the following dimensionful quantities, denoted by capital letters.
The half-width of the potential well is conveniently defined as $L=(\ell+1)a$,
while the conjugate momentum to the center-of-mass coordinate
is $Q=q/a$, and finally the continuum energy reads $E=(4-\o)/(2a^2)$.

To leading order as $a\to0$
(i.e., in the vicinity of the center of the Brillouin zone),
both dispersion relations~(\ref{wb}) and~(\ref{wf}) yield
\beq
E_k\approx\frac{Q^2}{4}+\eps_k,
\label{ec}
\eeq
where the $\eps_k$ are the energy levels of a free particle
in a potential well of width $2L$ in the appropriate sectors, i.e.,
\beq
\eps_k^\B=\left(\frac{(k+\half)\pi}{L}\right)^2,\qquad
\eps_k^\F=\left(\frac{k\pi}{L}\right)^2.
\eeq
The expression~(\ref{ec}) conforms
to what we expect from the continuum theory:
the total energy~$E_k$ of the compound system
is the sum of the kinetic energy $Q^2/4$ of the center-of-mass motion
and of the energy $\eps_k$ of a bound state in the relative coordinate.

The corrections to the leading-order result~(\ref{ec})
can be derived by recasting~(\ref{wb}) and~(\ref{wf}) in terms
of the dimensionful quantities introduced above, and expanding in powers of $a$.
The first correction thus obtained,
\beq
E_k=\frac{Q^2}{4}+\eps_k-\left(Q^4+24Q^2\eps_k+16\eps_k^2\right)\frac{a^2}{192}+\cdots,
\eeq
already demonstrates that the coupling of both degrees of freedom
by the lattice structure affects the energy spectrum in a non-trivial way.

\subsection{A case study: fermionic state with maximal distance $\ell=4$}

To close this section,
let us investigate in detail the dynamics
of a fermionic bound state with maximal distance $\ell=4$.
This is the smallest $\ell$ for which generic dynamical behavior is
observed.\footnote{For $\ell=3$, there exists a central anomalous front with zero speed.}
The front velocities~(\ref{vk}) read
\beqa
V_1^\F&=&-V_4^\F= 2\cos\frac{\pi}{5}=\tau=\frac{\sqrt{5}+1}{2},
\nonumber\\
V_2^\F&=&-V_3^\F=2\cos\frac{2\pi}{5}=\tin=\frac{\sqrt{5}-1}{2},
\label{vtt}
\eeqa
where $\tau$ is the golden mean and $\tin$ its reciprocal.

Suppose the two fermions are launched from sites 0 and~1 at time $t=0$.
Then $\p_{n,m}(0)=\delta_{n,0}\delta_{m,1}$ and
the initial mean value of the center-of-mass coordinate is
$\mean{n_\cm(0)}=\braopket{\p(0)}{n_\cm}{\p(0)}=1/2$.
We expect there will be two extremal ballistic fronts near $n=\pm\tau t$,
and two internal ones near $n=\pm\tin t$.

\begin{figure}[!ht]
\begin{center}
\includegraphics[angle=-90,width=.47\linewidth]{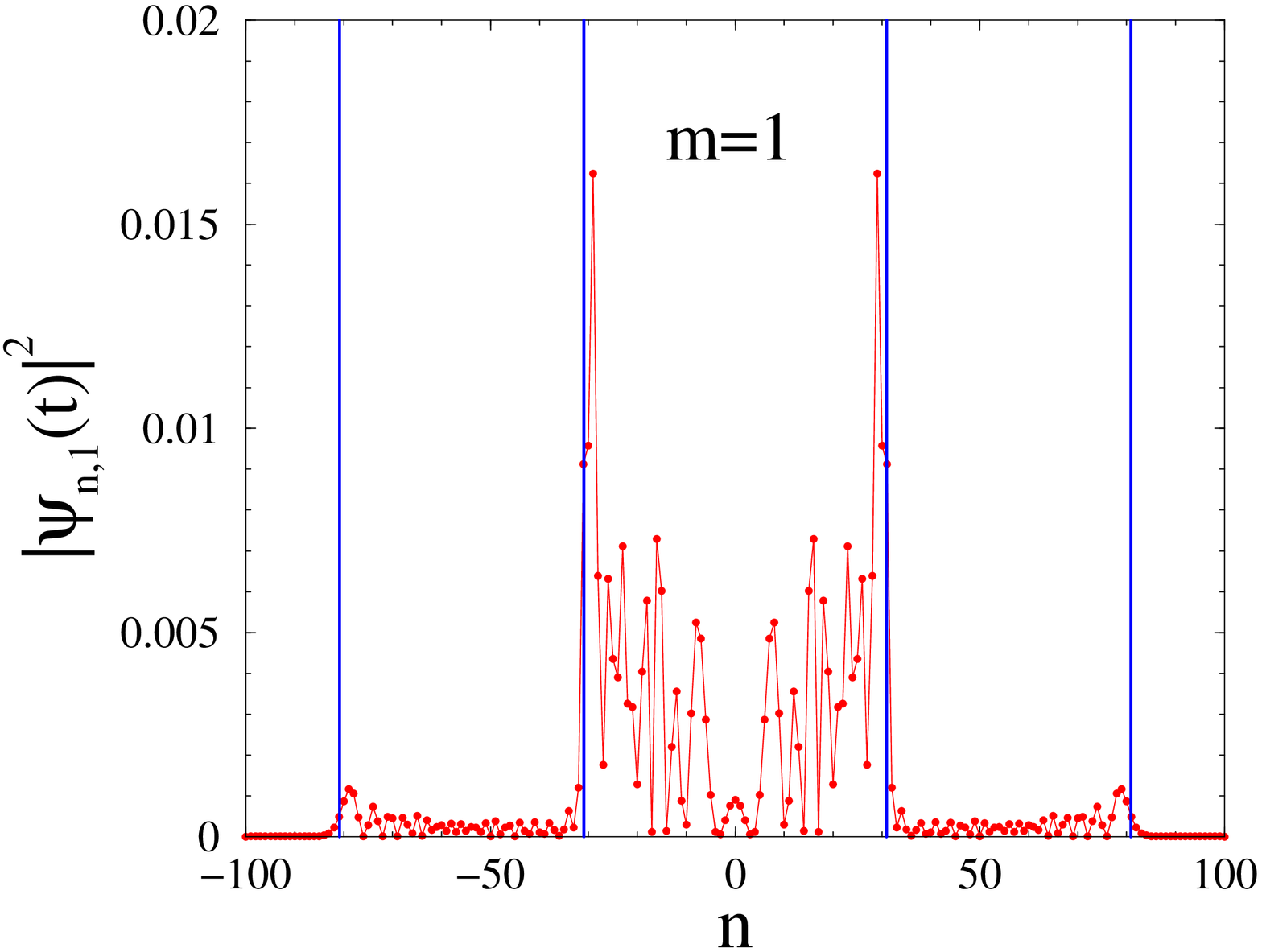}
\hskip 6pt
\includegraphics[angle=-90,width=.47\linewidth]{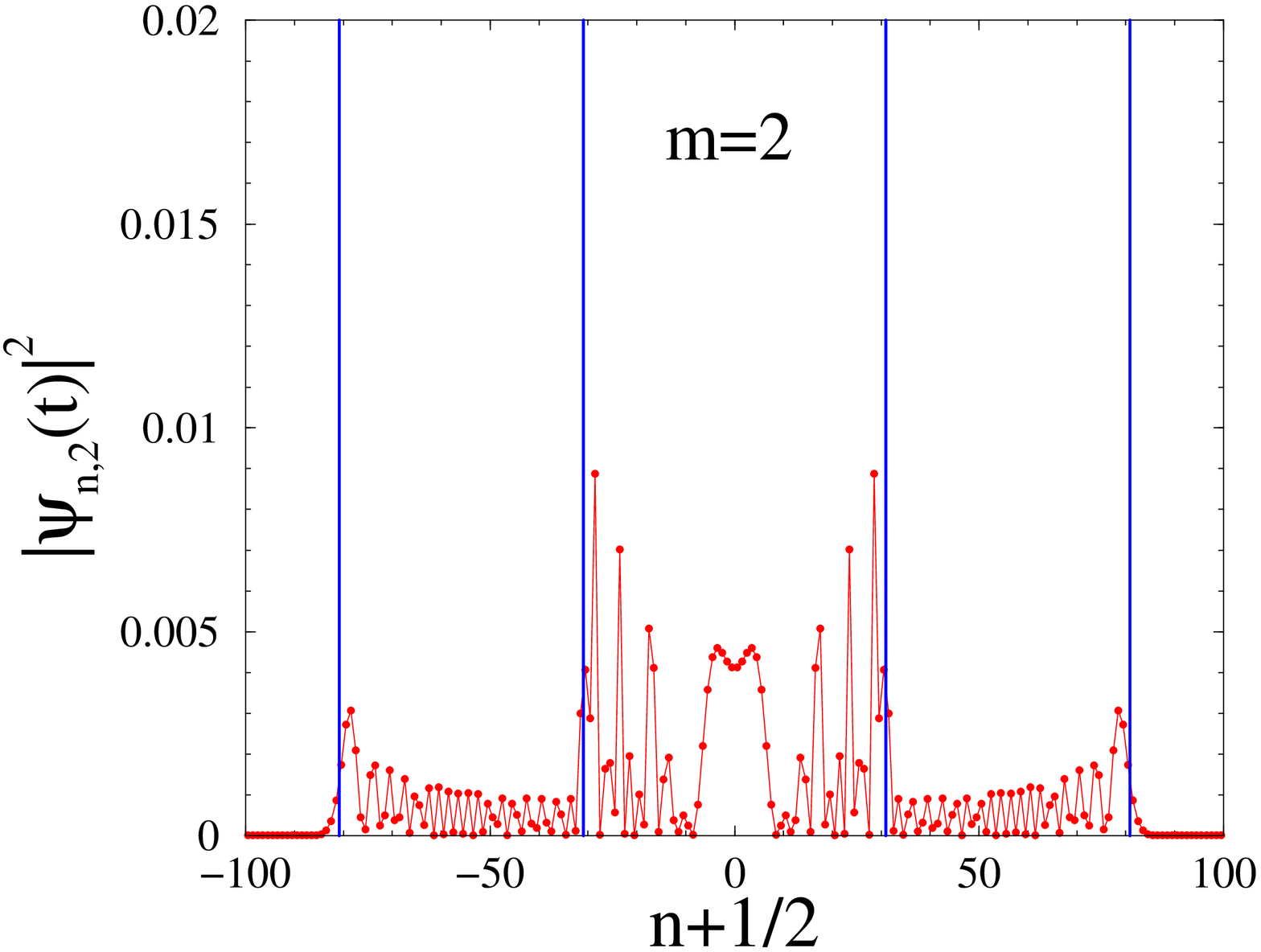}

\includegraphics[angle=-90,width=.47\linewidth]{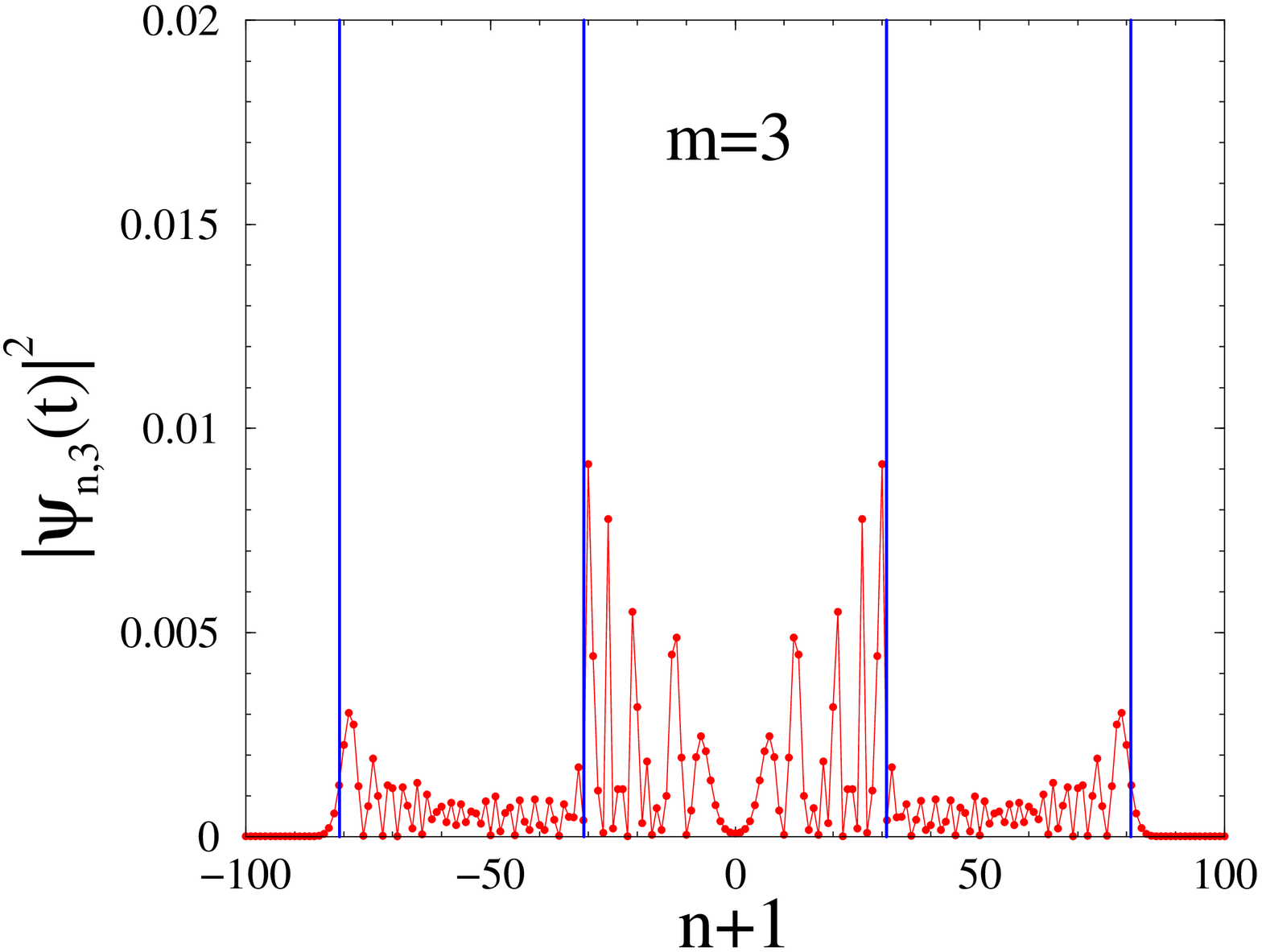}
\hskip 6pt
\includegraphics[angle=-90,width=.47\linewidth]{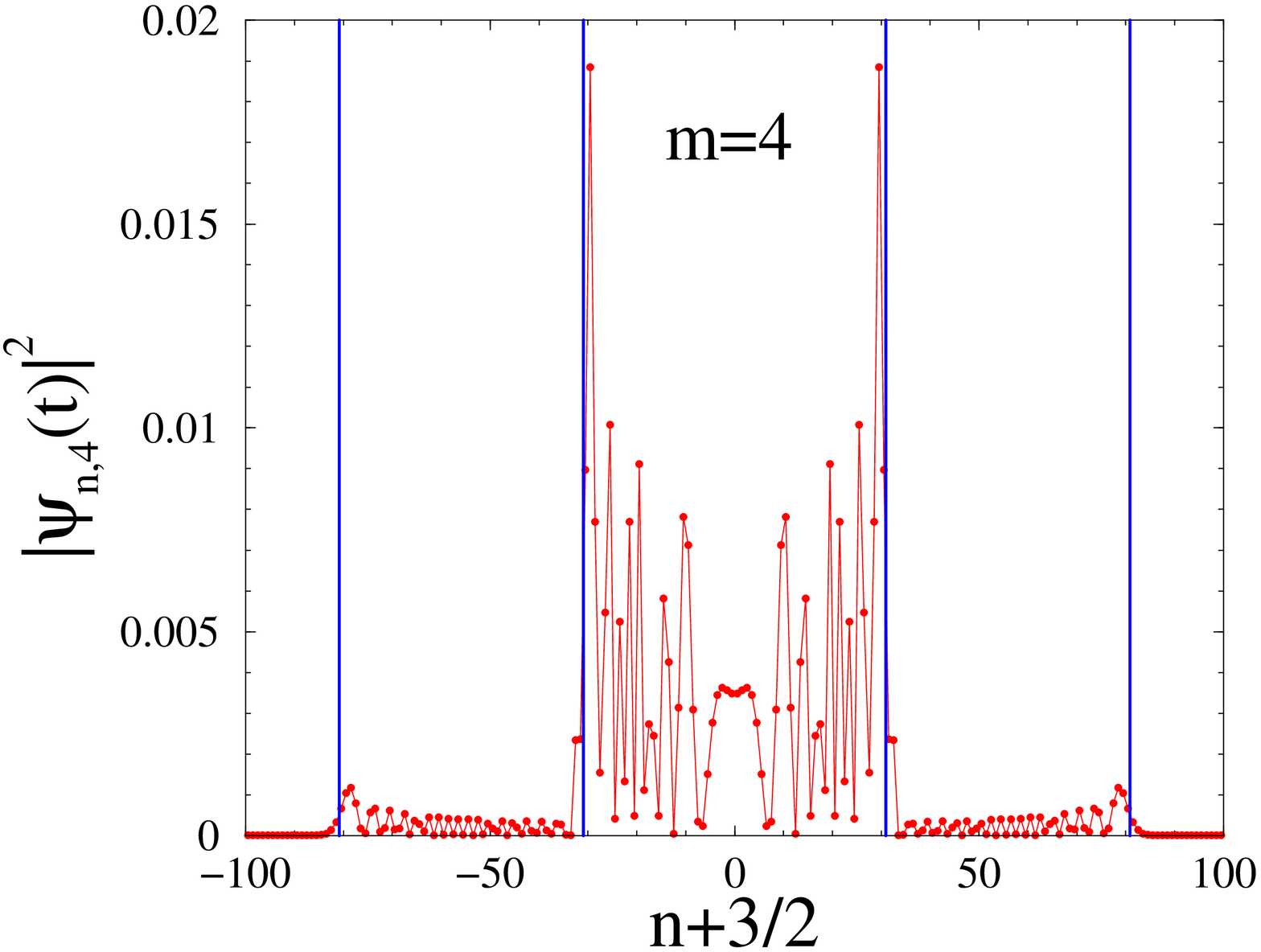}
\caption{\small
Probability profiles $\abs{\p_{n,m}(t)}^2$ ($m=1$ to 4)
of the fermionic bound state with maximal distance $\ell=4$ at time $t=50$,
with initial state $\p_{n,m}(0)=\delta_{n,0}\delta_{m,1}$,
against $n+(m-1)/2=n_\cm-\mean{n_\cm(0)}$.
Vertical blue lines: theoretical positions of the ballistic fronts.}
\label{f4}
\end{center}
\end{figure}

The most efficient way of solving the differential equations~(\ref{2psi}) with
prescribed initial values
consists in performing a spatial Fourier transform (with $q$ being conjugate to $n$)
and a temporal Laplace transform (with $s$ being conjugate to~$t$).
The Fourier-Laplace transforms thus defined obey the equations
\beqa
\ii s\h\p_1(q,s)=(\e^{-\ii q}+1)\h\p_2(q,s)+\ii,
\nonumber\\
\ii s\h\p_2(q,s)=(\e^{\ii q}+1)\h\p_1(q,s)+(\e^{-\ii q}+1)\h\p_3(q,s),
\nonumber\\
\ii s\h\p_3(q,s)=(\e^{\ii q}+1)\h\p_2(q,s)+(\e^{-\ii q}+1)\h\p_4(q,s),
\nonumber\\
\ii s\h\p_4(q,s)=(\e^{\ii q}+1)\h\p_3(q,s).
\eeqa
This linear system can be readily solved.
Introducing the notation
\beq
\g=2\cos\frac{q}{2},
\label{gdef}
\eeq
we obtain
\beqa
\h\p_1(q,s)=\frac{s(s^2+2\g^2)}{\Delta(q,s)},\qquad
\h\p_2(q,s)=-\ii\frac{(\e^{\ii q}+1)(s^2+\g^2)}{\Delta(q,s)},
\nonumber\\
\h\p_3(q,s)=-\frac{(\e^{\ii q}+1)^2s}{\Delta(q,s)},\qquad
\h\p_4(q,s)=\ii\frac{(\e^{\ii q}+1)^3}{\Delta(q,s)},
\label{pfl}
\eeqa
with
\beq
\Delta(q,s)=(s-\ii\g\tau)(s-\ii\g\tin)(s+\ii\g\tau)(s+\ii\g\tin).
\eeq
The inverse Laplace transform of~(\ref{pfl}) can be taken first.
We have~e.g.
\beq
\h\p_1(q,s)=\frac{1}{2\sqrt5}\left(\frac{\tin}{s-\ii\g\tau}+\frac{\tin}{s+\ii\g\tau}
+\frac{\tau}{s-\ii\g\tin}+\frac{\tau}{s+\ii\g\tin}\right),
\eeq
and so
\beq
\h\p_1(q,t)=\frac{1}{2\sqrt5}\left(\tin\e^{\ii\g\tau t}+\tin\e^{-\ii\g\tau t}
+\tau\e^{\ii\g\tin t}+\tau\e^{-\ii\g\tin t}\right).
\eeq
The inverse Fourier transform can then be taken using (see~(\ref{pint}))
\beqa
\int_0^{2\pi}\frac{\d q}{2\pi}\,\e^{\ii(nq+x\cos(q/2))}=(-1)^nJ_{2n}(x),
\nonumber\\
\int_0^{2\pi}\frac{\d q}{2\pi}\,\e^{\ii((n+1/2)q+x\cos(q/2))}=\ii(-1)^nJ_{2n+1}(x).
\label{j2n}
\eeqa
We thus obtain
\beqa
\p_{n,1}(t)&=&\frac{(-1)^n}{\sqrt5}
\bigl(\tin J_{2n}(2\tau t)+\tau J_{2n}(2\tin t)\bigr),
\nonumber\\
\p_{n,2}(t)&=&\ii\frac{(-1)^{n+1}}{\sqrt5}
\bigl(J_{2n+1}(2\tau t)+J_{2n+1}(2\tin t)\bigr),
\nonumber\\
\p_{n,3}(t)&=&\frac{(-1)^{n+1}}{\sqrt5}
\bigl(J_{2n+2}(2\tau t)-J_{2n+2}(2\tin t)\bigr),
\nonumber\\
\p_{n,4}(t)&=&\ii\frac{(-1)^n}{\sqrt5}
\bigl(\tin J_{2n+3}(2\tau t)-\tau J_{2n+3}(2\tin t)\bigr).
\label{p4}
\eeqa
Figure~\ref{f4} shows the four probability profiles $\abs{\p_{n,m}(t)}^2$
thus obtained at time $t=50$.
Abscissas are shifted from $n$ to $n+(m-1)/2=n_\cm-\mean{n_\cm(0)}$,
in such a way that the plotted profiles are exactly symmetric.
The four profiles exhibit the same global features,
although they differ in their detailed structure.
As predicted,
they exhibit the same ballistic fronts, two external ones at $n\approx\pm\tau t$
and two internal ones at $n\approx\pm\tin t$ (see~(\ref{vtt})).
The probabilities are larger within the internal fronts,
and considerably smaller in the wings of the allowed zone,
i.e., between the internal and the external fronts.
This phenomenon is quite generic; it could already be observed
in the left panel of figure~\ref{pext}.

Let us now turn to the internal structure of the fermionic bound state.
The distance between both particles takes the value $m=1,\dots,4$ with probability
\beq
P_m(t)=\sum_n\abs{\p_{n,m}(t)}^2.
\eeq
These probabilities can be evaluated from~(\ref{p4}) using identities
\beqa
\sum_nJ_{2n}(x)J_{2n}(y)=\frac{J_0(x-y)+J_0(x+y)}{2},
\nonumber\\
\sum_nJ_{2n+1}(x)J_{2n+1}(y)=\frac{J_0(x-y)-J_0(x+y)}{2},
\eeqa
which can be derived from the integral representations~(\ref{j2n}).
Quadratic identities of this kind have their roots in the connection
between special functions and representation theory~\cite{vilenkin}.
We thus obtain
\beqa
P_1(t)=\frac{1}{10}\bigl(3+\tin^2J_0(4\tau t)+\tau^2J_0(4\tin t)
+2J_0(2t)+2J_0(2\sqrt5t)\bigr),
\nonumber\\
P_2(t)=\frac{1}{10}\bigl(2-J_0(4\tau t)-J_0(4\tin t)
+2J_0(2t)-2J_0(2\sqrt5t)\bigr),
\nonumber\\
P_3(t)=\frac{1}{10}\bigl(2+J_0(4\tau t)+J_0(4\tin t)
-2J_0(2t)-2J_0(2\sqrt5t)\bigr),
\nonumber\\
P_4(t)=\frac{1}{10}\bigl(3-\tin^2J_0(4\tau t)-\tau^2J_0(4\tin t)
-2J_0(2t)+2J_0(2\sqrt5t)\bigr).
\label{p4int}
\eeqa
These probabilities sum up to unity, as should be.
They are even functions of $t$ whose power series have rational coefficients.
At short times we have
\beqa
P_1(t)=1-2t^2+\frac{5t^4}{2}-\frac{35t^6}{18}\cdots,
\nonumber\\
P_2(t)=2t^2-4t^4+\frac{35t^6}{9}+\cdots,
\nonumber\\
P_3(t)=\frac{3t^4}{2}-\frac{5t^6}{2}+\cdots,
\nonumber\\
P_4(t)=\frac{5t^6}{9}+\cdots
\eeqa
In the long-time regime, these probabilities reach the stationary values
\beq
P_1=P_4=\frac{3}{10},\qquad P_2=P_3=\frac{1}{5}.
\label{pstat}
\eeq
These numbers agree with the result~(\ref{q4}) derived in~\ref{app}.
In the present situation,
it is indeed legitimate to study the stationary properties of the internal state per se,
without referring to the ballistic dynamics of the center-of-mass coordinate,
because the compound system has a basis of factorized eigenstates~(\ref{pw}).
This feature is by no means general.

\begin{figure}[!ht]
\begin{center}
\includegraphics[angle=-90,width=.5\linewidth]{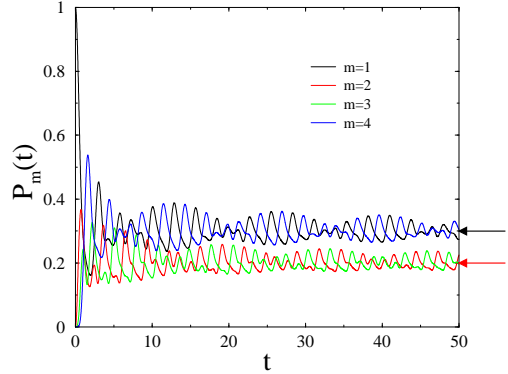}
\caption{\small
Probabilities $P_m(t)$ describing the internal structure of the fermionic bound state
(see text) against time $t$.
Arrows: stationary values~(\ref{pstat}).}
\label{p4m}
\end{center}
\end{figure}

Figure~\ref{p4m} shows a plot of the probabilities $P_m(t)$ against time $t$.
The maximal time $t=50$ corresponds to the profiles shown in figure~\ref{f4}.
The stationary values~(\ref{pstat}) (arrows) are reached
after a complex pattern of damped oscillations.
The envelope of this transient oscillatory behavior falls off very slowly as $t^{-1/2}$,
while the oscillations themselves follow a quasiperiodic pattern,
characterized by the two incommensurate frequencies $2\tau$ and $2\tin$.
All the frequencies entering~(\ref{p4int}) are indeed integer linear combinations
of the latter frequencies.
In a generic situation, there will be as many incommensurate frequencies
as there are positive ballistic velocities $V_k^\B$ or $V_k^\F$ (see~(\ref{vk})).

\begin{figure}[!ht]
\begin{center}
\includegraphics[angle=-90,width=.5\linewidth]{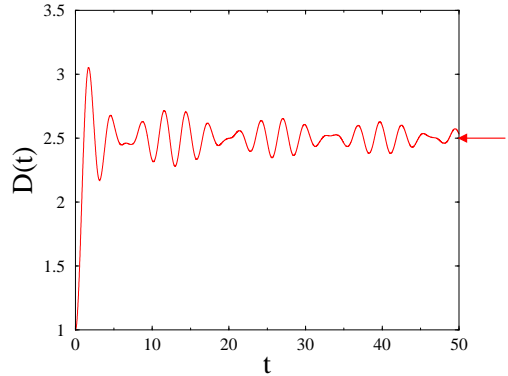}
\caption{\small
Plot of the mean internal size $D(t)$ of the fermionic bound state against time $t$.
Arrow: stationary value ($D=5/2$).}
\label{d4}
\end{center}
\end{figure}

The mean internal size of the bound state,
\beq
D(t)=\sum_{m=1}^4 m P_m(t),
\eeq
can be evaluated from~(\ref{p4int}) to read
\beq
D(t)=\frac{1}{10}\bigl(25-(2-3\tin)J_0(4\tau t)-(2+3\tau)J_0(4\tin t)-8J_0(2t)\bigr).
\eeq
This quantity is plotted in figure~\ref{d4}.
It starts from $D(0)=1$ and reaches the stationary value $D=5/2$,
again after a complex pattern of quasiperiodic oscillations dying off as $t^{-1/2}$.

\section{Two-body bound states: smooth confining potential}
\label{ssoft}

\subsection{Generalities}

We now turn to the quantum walk
performed by a bosonic or fermionic bound state of two identical particles
generated by a smooth confining potential $W_m$.
The latter
is assumed to be an even function of the relative coordinate $m$ between both particles,
such that $W_m\to+\infty$ at large distances ($\abs{m}\to\infty$).
With the notations of section~\ref{shard},
the time-dependent wavefunction $\p_{n,m}(t)$ obeys the equation
\beqa
\ii\,\dif{\p_{n,m}(t)}{t}=W_m\p_{n,m}(t)&+&\p_{n,m-1}(t)+\p_{n+1,m-1}(t)
\nonumber\\
&+&\p_{n-1,m+1}(t)+\p_{n,m+1}(t).
\label{2vdif}
\eeqa

Let us look for a basis of solutions to~(\ref{2vdif}) of the form (see~(\ref{pw}))
\beq
\p_{n,m}(t)=\e^{\ii(n_\cm q-\o t)}\phi_m,
\label{phi}
\eeq
where the momentum $q$ is conjugate to the center-of-mass coordinate~(\ref{ncm}).
The internal wavefunction $\phi_m$ obeys
\beq
\o\phi_m=\g(\phi_{m-1}+\phi_{m+1})+W_m\phi_m,
\label{ephi}
\eeq
with the same dispersive (i.e.,~$q$-dependent) hopping amplitude
\beq
\g=2\cos\frac{q}{2}
\eeq
as before (see~(\ref{gdef})).
We shall also use the shorthand notation
\beq
\chi_m=(-1)^m\phi_m,
\eeq
for the staggered wavefunction, which obeys
\beq
\o\chi_m=-\g(\chi_{m-1}+\chi_{m+1})+W_m\chi_m,
\label{echi}
\eeq
i.e.,~(\ref{ephi}) with the sign of $\g$ reversed.

The above formalism applies to an arbitrary confining potential.
Bosonic and fermionic bound states are described by wavefunctions $\phi_m$
which are respectively even and odd functions of~$m$,
such that $\phi_m\to0$ at large distances.
Such bound-state wavefunctions only exist for discrete sequences of dispersive
frequencies $\o_k^\B(q)$ and $\o_k^\F(q)$.
Hereafter we investigate in detail the case of potentials growing as a power of distance,
either linearly (section~\ref{soft1}), quadratically (section~\ref{soft2}),
or as an arbitrary power (section~\ref{softa}).

\subsection{Linear confining potential}
\label{soft1}

Our first example is the linear confining potential
\beq
W_m=g\abs{m},
\eeq
where the amplitude $g$ is a positive constant.
The regime of most physical interest corresponds to small $g$,
so that bound states have a large size and potentially a rich internal structure.

\begin{figure}[!ht]
\begin{center}
\includegraphics[angle=-90,width=.5\linewidth]{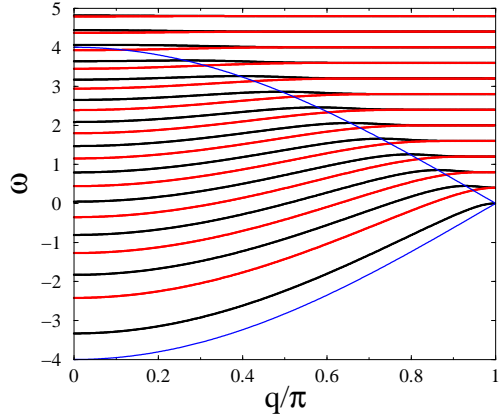}
\caption{\small
Plot of the bosonic and fermionic spectra
for the linear confining potential with $g=0.4$, against $q/\pi$.
Black: bosonic frequencies $\o_k^\B(q)$.
Red: fermionic frequencies $\o_k^\F(q)$.
Blue: band boundaries ($\o=\pm2\g=\pm4\cos(q/2)$).}
\label{wlin}
\end{center}
\end{figure}

Figure~\ref{wlin} shows the bosonic and fermionic spectra for $g=0.4$.
These spectra exhibit two very distinct regions.
Within the band delimited by the blue curves ($\o=\pm2\g=\pm4\cos(q/2)$),
bosonic and fermionic branches are well separated from each other.
They alternate and are strongly dispersive.
Above the band, the spectrum consists of an infinite array
of approximately equally spaced and non-dispersive branches.
These branches appear as red,
as bosonic and fermionic branches are superimposed to a very high accuracy.

In order to analyze the above observations,
we use~(\ref{echi}) for positive values of the relative coordinate $m$, i.e.,
\beq
\o\chi_m=-\g(\chi_{m-1}+\chi_{m+1})+gm\chi_m\qquad(m\ge0).
\label{eladder}
\eeq

Let us first forget about the constraint $m\ge0$
and extend~(\ref{eladder}) to all values of~$m$.
We are thus facing a tight-binding equation
for a charged particle in a uniform electric field,
the amplitude $g$ giving the strength of the field in reduced units.
The corresponding spectrum is a Wannier-Stark ladder~\cite{wbook,wrev}:
it consists of the infinite sequence of equally spaced frequencies,
\beq
\o_k=gk,
\label{ws}
\eeq
where $k$ runs over the integers.
The Wannier-Stark levels are non-dispersive,
as~(\ref{ws}) holds irrespective of the hopping amplitude $\g$.
The normalized eigenfunction corresponding to $\o_k$ reads
\beq
\chi_m=J_{m-k}(2\g/g),
\eeq
where the $J_m$ are the Bessel functions.
This eigenfunction is strongly localized around $m=k$.
We have in particular, for large positive $k$,
\beq
\abs{\chi_0}\approx\frac{(\g/g)^k}{k!}\sim\left(\frac{\e\g}{gk}\right)^k.
\label{chi0}
\eeq
This estimate shows that the Wannier-Stark eigenfunctions
hardly feel the boundary condition on~(\ref{eladder}) at $m=0$,
and therefore remain essentially unperturbed,
as soon as the frequency $\o_k=gk$ exceeds a few times the bandwidth $\g$.
In other words, only finitely many lowest Wannier-Stark states,
in a number of the order of $\g/g$, are affected by the boundary condition
which distinguishes between fermions and bosons.
This explains the main observations made on figure~\ref{wlin}.

A more quantitative analysis of the problem goes as follows.
The general solution to~(\ref{eladder}) falling off as $m\to+\infty$ reads
\beq
\chi_m=J_{m-\o/g}(2\g/g)\qquad(m\ge-1),
\eeq
where the $J_\nu(z)$ are the Bessel functions with index $\nu$,
which obey the identities
\beqa
J_{\nu-1}(z)+J_{\nu+1}(z)&=&\frac{2\nu}{z}\,J_\nu(z),
\nonumber\\
J_{\nu-1}(z)-J_{\nu+1}(z)&=&2J'_\nu(z),
\label{jiden}
\eeqa
where the prime denotes a derivative with respect to $z$.
The bosonic spectrum is given by $\chi_1=\chi_{-1}$,
while for the fermionic spectrum $\chi_0=0$.
We thus arrive at the exact quantization formulas
\beq
J'_{-\o^\B/g}(2\g/g)=0,\qquad
J_{-\o^\F/g}(2\g/g)=0.
\label{qlin}
\eeq

The leading correction to the Wannier-Stark spectrum~(\ref{ws}) at large positive $k$
can be readily derived from~(\ref{qlin}).
Skipping details, we obtain a symmetric splitting of the form
\beq
\o_k^\B\approx g(k+\eps_k),\qquad\o_k^\F\approx g(k-\eps_k),
\label{bf}
\eeq
with
\beq
\eps_k\approx\frac{(\g/g)^{2k}}{k!(k-1)!}\approx k\chi_0^2,
\label{bfs}
\eeq
where $\chi_0$ is the Wannier-Stark wavefunction at the origin (see~(\ref{chi0})).

Our next goal is to investigate the velocities
characterizing the ballistic spreading of a bosonic or fermionic wavefunction
in the center-of-mass coordinate $n_\cm$, for a generic initial state.
These velocities are given by
\beq
V^\B=\comport{\max}{k,q}\abs{\o_k'^\B(q)},\qquad
V^\F=\comport{\max}{k,q}\abs{\o_k'^\F(q)},
\eeq
where primes denote derivatives,
and maxima are taken over all branches of the spectrum and over all~$q$.
It is clear from figure~\ref{wlin} that these maxima are reached for the lowest
branch of each spectrum.
Furthermore, if the amplitude $g$ is small,
so that the bound states have a large internal size,
we anticipate that the maximal velocities will be close to the free value 2
and correspond to momenta close to the zone boundary ($q\to\pi$).

To substantiate these expectations we note that whenever $g$ is small,
typical wavefunctions vary slowly with $m$, so one can employ a continuum description.
Setting
\beq
\o=-2\g+\delta=-4\cos\frac{q}{2}+\delta,
\label{odelta}
\eeq
and expanding in~(\ref{eladder}) differences in terms of derivatives, we obtain
\beq
\g\frac{\d^2\chi}{\d m^2}\approx(gm-\delta)\chi.
\label{dif2}
\eeq
Then, setting
\beq
x=\lambda(gm-\delta),\qquad\lambda=(g^2\g)^{-1/3},
\eeq
Equation~(\ref{dif2}) becomes the Airy equation
\beq
\frac{\d^2\chi}{\d x^2}=x\chi,
\eeq
whose solution decaying as $x\to+\infty$ is $\chi(x)=\Ai(x)$, the Airy function.

\begin{itemize}

\item {\it Bosonic states}
obey $\chi_1=\chi_{-1}$, hence $\d\chi/\d m=0$ for $m=0$,
and so $x_0=-\lambda\delta$ obeys $\Ai'(x_0)=0$.
The lowest branch of the bosonic spectrum therefore reads
\beq
\delta^\B\approx\eta_1(g^2\g)^{1/3},
\label{delb}
\eeq
where $\eta_1=1.018792\dots$ is the opposite of the first zero of $\Ai'(x)$.

Setting $q=\pi-\eps$,
the lowest branch can be expanded as
\beq
\o=-4\sin\frac{\eps}{2}+\delta=-2\eps+\frac{\eps^3}{12}+\cdots+\delta.
\label{oexpand}
\eeq
Using the expression~(\ref{delb}) of $\delta$,
we can show that the maximal group velocity reads
\beq
V^\B\approx2-C^\B\,g^{1/2},
\label{vblin}
\eeq
\beq
C^\B=\left(\frac{4\eta_1}{9}\right)^{3/4}=0.551985\dots,
\eeq
and is reached for
\beq
\eps\approx(C^\B)^{1/2}\,g^{1/4}.
\eeq

\item {\it Fermionic states} obey $\chi_0=0$, and
so $x_0=-\lambda\delta$ obeys $\Ai(x_0)=0$.
The lowest branch of the fermionic spectrum therefore reads
\beq
\delta^\F\approx\xi_1(g^2\g)^{1/3},
\label{delf}
\eeq
where $\xi_1=2.338107\dots$ is the opposite of the first zero of $\Ai(x)$.
The maximal group velocity now reads
\beq
V^\F\approx2-C^\F\,g^{1/2},
\label{vflin}
\eeq
\beq
C^\F=\left(\frac{4\xi_1}{9}\right)^{3/4}=1.029227\dots,
\eeq
and it is reached for
\beq
\eps\approx(C^\F)^{1/2}\,g^{1/4}.
\eeq

\end{itemize}

\begin{figure}[!ht]
\begin{center}
\includegraphics[angle=-90,width=.5\linewidth]{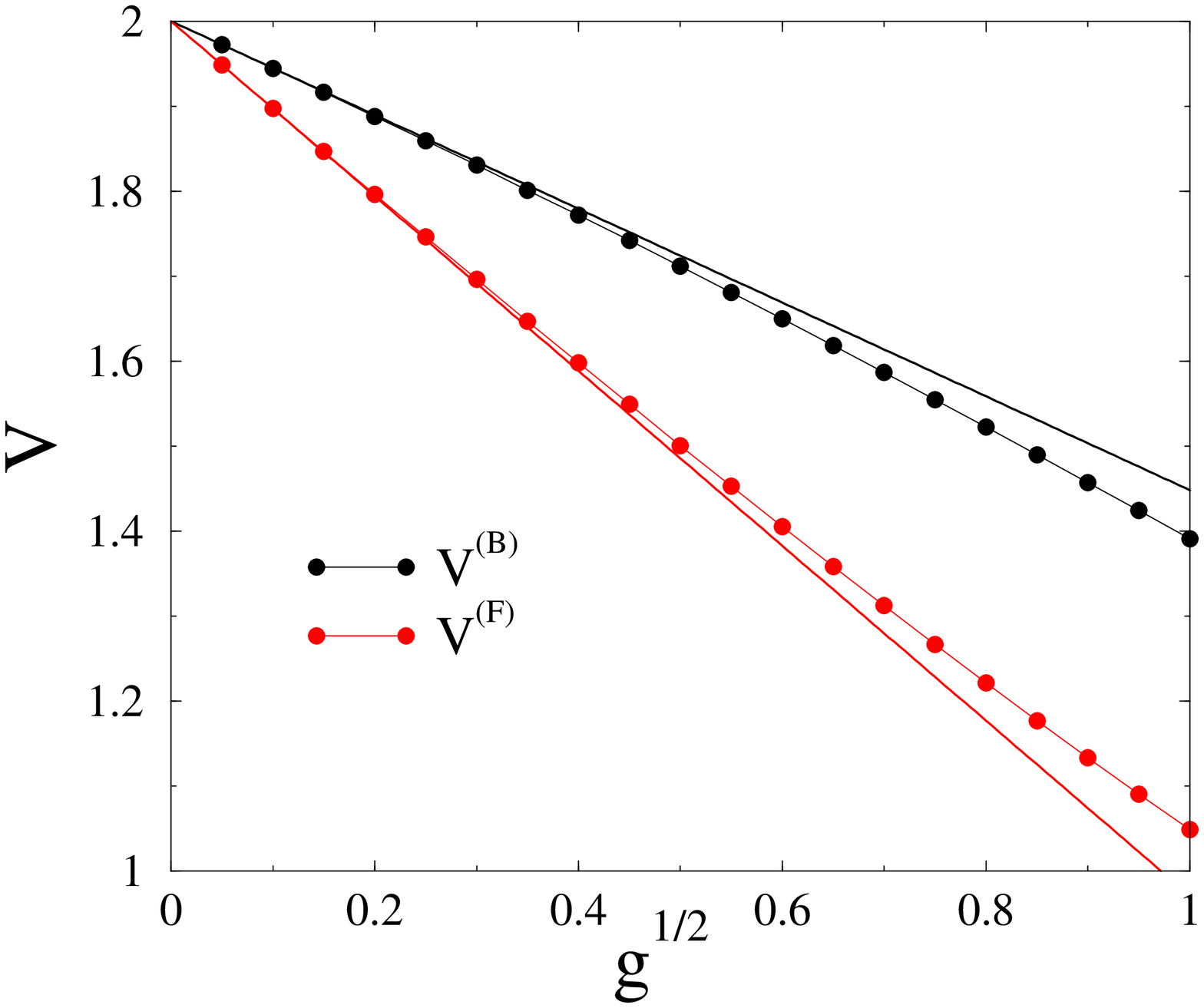}
\caption{\small
Plot of the bosonic and fermionic maximal velocities $V^\B$ (black) and $V^\F$ (red),
characterizing ballistic spreading
in a linear confining potential, against $g^{1/2}$.
Straight lines: scaling results~(\ref{vblin}) and~(\ref{vflin}) at small $g$.}
\label{vlin}
\end{center}
\end{figure}

The above scaling results valid in the $g\ll 1$ regime can be alternatively derived
by analyzing the exact quantization formulas~(\ref{qlin})
in the transition region (see~(\ref{jtrans})).
We have in particular
\beq
\frac{C^\F}{C^\B}=\left(\frac{\xi_1}{\eta_1}\right)^{3/4}=1.864592\dots
\label{cfb1}
\eeq
Figure~\ref{vlin} shows plots of the maximal velocities $V^\B$ and $V^\F$
against~$g^{1/2}$.
The scaling results~(\ref{vblin}) and~(\ref{vflin}) at small $g$ (straight lines)
provide a good description of the velocities for moderate values of $g$.

\subsection{Quadratic confining potential}
\label{soft2}

Our second example is the quadratic confining potential
\beq
W_m=gm^2,
\eeq
where $g$ is again a positive constant.
Equation~(\ref{echi}) reads
\beq
\o\chi_m=-\g(\chi_{m-1}+\chi_{m+1})+gm^2\chi_m.
\label{qchi}
\eeq
The Fourier transform $\h\chi(p)$ therefore obeys
\beq
g\frac{\d^2\h\chi}{\d p^2}=-(\o+2\g\cos p)\h\chi.
\eeq
The latter differential equation is known as the Mathieu equation~\cite{abramowitz}.
The body of knowledge on the latter equation is however of little use
for the present purpose.
Hereafter we therefore use general techniques
which could be applied to any confining potential.

\begin{figure}[!ht]
\begin{center}
\includegraphics[angle=-90,width=.5\linewidth]{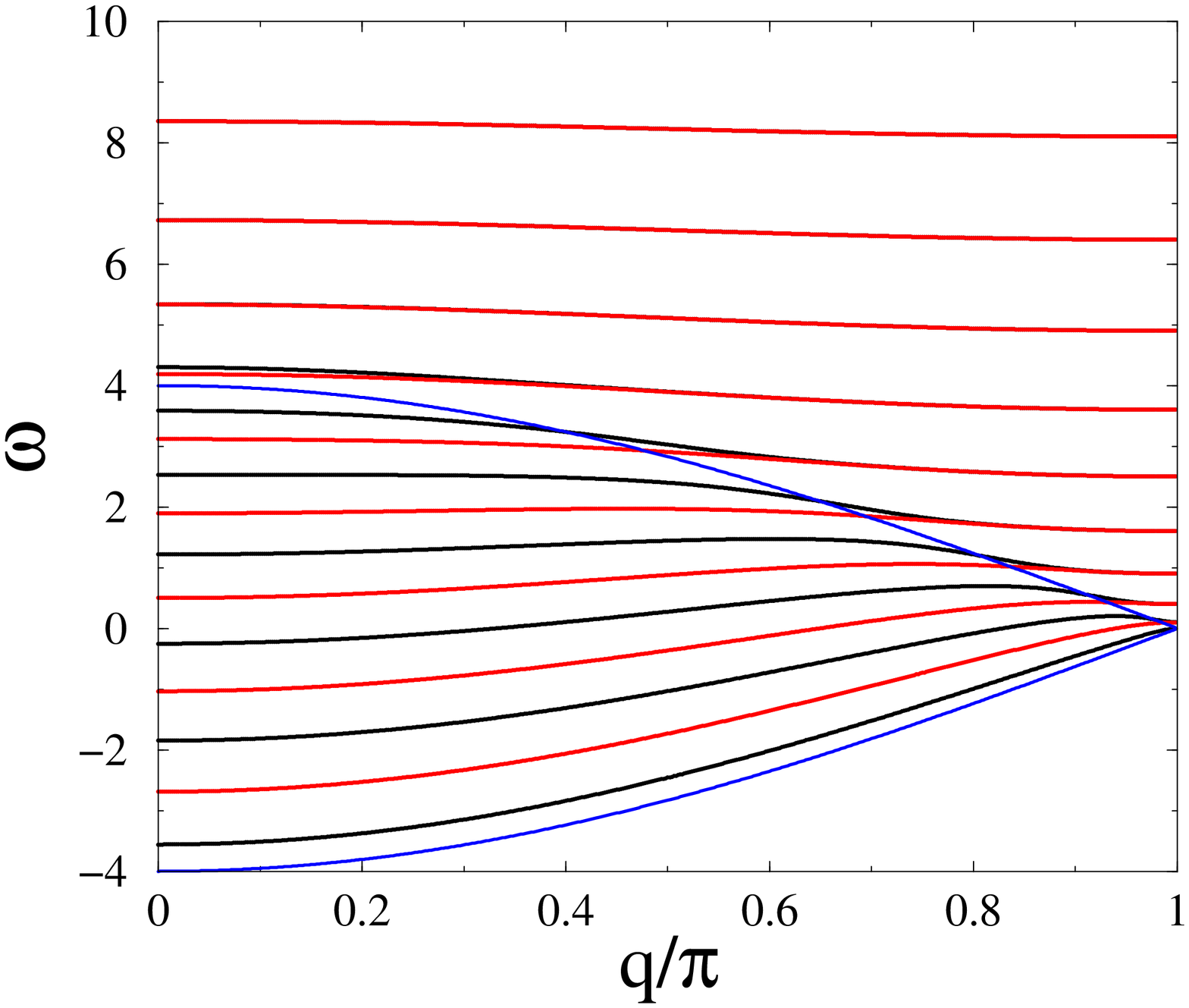}
\caption{\small
Plot of the bosonic and fermionic spectra
for the quadratic confining potential with $g=0.1$, against $q/\pi$.
Black: bosonic frequencies $\o_k^\B(q)$.
Red: fermionic frequencies $\o_k^\F(q)$.
Blue: band boundaries ($\o=\pm2\g=\pm4\cos(q/2)$).}
\label{wquad}
\end{center}
\end{figure}

Figure~\ref{wquad} shows the bosonic and fermionic spectra for $g=0.1$.
These spectra are qualitatively similar to those shown in figure~\ref{wlin}.
The existence of two very distinct regions,
with strongly dispersive modes within the band
and weakly dispersive branches above the band,
is indeed a common feature of all confining potentials.

Let us first consider the weakly dispersive branches above the band.
Right at $\g=0$
(this corresponds to $q=\pi$, i.e., to the right end of figure~\ref{wquad}),
the eigenstates are strictly localized at specific sites:
we have $\chi_m=\delta_{m,k}$ and $\o_k=W_k=gk^2$.
If $\g$ is small and/or $k$ is large,
the wavefunction $\chi_{k\pm1}$ at neighboring sites
is proportional to the ratio $\g/k$.
We thus obtain the more refined estimate
\beq
\o_k\approx gk^2+\frac{2\g^2}{(4k^2-1)g}.
\eeq
High branches are therefore weakly dispersive, as their bandwidth scales as $1/(gk^2)$.
With a linear confining potential,
the Wannier-Stark states were not dispersive at all.

In the whole region above the band,
the splitting between even (bosonic) and odd (fermionic) frequencies
can hardly be observed.
This splitting can be estimated as follows.
First, it is expected on general grounds to scale as $\chi_0^2$ (see~(\ref{bfs})).
Furthermore, $\chi_0$ is exponentially small whenever $\g$ is small and/or $k$ is large.
In this regime,~(\ref{qchi}) indeed simplifies to
$g(k^2-m^2)\chi_m\approx-\g\chi_{m+1}$ for $0<m<k$.
We thus obtain the estimate
\beq
\abs{\chi_0}\approx\frac{2(\g/g)^k}{(2k)!}\sim\left(\frac{\e^2\g}{4gk^2}\right)^k.
\label{qchi0}
\eeq

The velocities $V^\B$ and $V^\F$
characterizing the ballistic spreading of an initially localized wavefunction
again correspond to the lowest branches of the spectra.
In the regime of most interest where $g$ is small,
these velocities can be calculated by means of a continuum description.
With the notation~(\ref{odelta}), we obtain
\beq
\g\frac{\d^2\chi}{\d m^2}\approx(gm^2-\delta)\chi.
\label{qdif2}
\eeq
Setting
\beq
m=(g/\g)^{1/4}x,\qquad\delta=(g\g)^{1/2}\E,
\eeq
Equation~(\ref{qdif2}) becomes the usual Schr\"odinger equation for a harmonic oscillator,
\beq
-\frac{\d^2\chi}{\d x^2}+x^2\chi=\E\chi,
\eeq
whose eigenvalues are $\E_\nu=2\nu+1$ ($\nu=0,1,\dots$).

For the bosonic spectrum, the lowest branch corresponds to $\E_0=1$, and so
\beq
\delta^\B\approx(g\g)^{1/2}.
\eeq
Using~(\ref{oexpand}), we obtain
\beq
V^\B\approx2-C^\B\,g^{2/5},
\label{vbquad}
\eeq
\beq
C^\B=\frac{5}{2^{14/5}}=0.717936\dots
\eeq

For the fermionic spectrum, the lowest branch corresponds to $\E_1=3$, and so
\beq
\delta^\F\approx3(g\g)^{1/2},
\eeq
\beq
V^\F\approx2-C^\F\,g^{2/5},
\label{vfquad}
\eeq
\beq
C^\F=\frac{3^{4/5}\,5}{2^{14/5}}=1.728952\dots
\eeq

The ratio of the amplitudes is
\beq
\frac{C^\F}{C^\B}=3^{4/5}=2.408224\dots
\label{cfb2}
\eeq

\begin{figure}[!ht]
\begin{center}
\includegraphics[angle=-90,width=.5\linewidth]{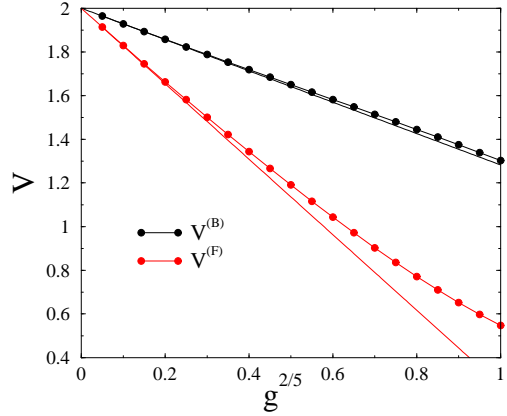}
\caption{\small
Plot of the bosonic and fermionic maximal velocities $V^\B$ (black) and $V^\F$ (red),
characterizing ballistic spreading
in a quadratic confining potential, against $g^{2/5}$.
Straight lines: scaling results~(\ref{vbquad}) and~(\ref{vfquad}) at small $g$.}
\label{vquad}
\end{center}
\end{figure}

Figure~\ref{vquad} shows plots of the maximal velocities $V^\B$ and $V^\F$
against~$g^{2/5}$.
Here again, the scaling results~(\ref{vbquad}) and~(\ref{vfquad}) at small $g$
(straight lines) provide a good description of the velocities for moderate values of $g$.

\subsection{Confining potential with an arbitrary exponent}
\label{softa}

We now consider the case of a confining potential of the form
\beq
W_m=g\abs{m}^\a,
\label{va}
\eeq
with an arbitrary exponent $\a>0$.

Let us focus on the maximal velocities
$V^\B$ and $V^\F$ characterizing the ballistic spreading of a localized initial state.
In the most interesting regime when $g$ is small,
these velocities can be calculated by means of a continuum description.
With the notation~(\ref{odelta}), we obtain
\beq
\g\frac{\d^2\chi}{\d m^2}\approx(g\abs{m}^\a-\delta)\chi.
\label{qdifa}
\eeq
Setting
\beq
m=(g/\g)^{1/(\a+2)}x,\qquad\delta=(g^2\g^\a)^{1/(\a+2)}\E,
\eeq
Equation~(\ref{qdifa}) becomes the continuous Schr\"odinger equation
\beq
-\frac{\d^2\chi}{\d x^2}+\abs{x}^\a\chi=\E\chi,
\label{scha}
\eeq
whose eigenvalues $\E_\nu$ ($\nu=0,1,\dots$) are not known analytically in general.

For the bosonic spectrum, the lowest branch corresponds to $\E_0$, and so
\beq
\delta^\B\approx(g^2\g^\a)^{1/(\a+2)}\E_0.
\eeq
Using again~(\ref{oexpand}), we obtain
\beq
V^\B\approx2-C^\B\,g^{2/(\a+3)},
\label{vba}
\eeq
\beq
C^\B=\frac{\a+3}{2^{2/(\a+3)}}\left(\frac{\a\E_0}{(\a+2)^2}\right)^{(\a+2)/(\a+3)}.
\label{cba}
\eeq

For the fermionic spectrum, the lowest branch corresponds to $\E_1$, and so
\beq
\delta^\F\approx(g^2\g^\a)^{1/(\a+2)}\E_1,
\eeq
\beq
V^\F\approx2-C^\F\,g^{2/(\a+3)},
\label{vfa}
\eeq
\beq
C^\F=\frac{\a+3}{2^{2/(\a+3)}}\left(\frac{\a\E_1}{(\a+2)^2}\right)^{(\a+2)/(\a+3)}.
\label{cfa}
\eeq

The ratio of the amplitudes is
\beq
\frac{C^\F}{C^\B}=\left(\frac{\E_1}{\E_0}\right)^{(\a+2)/(\a+3)}.
\label{cfcba}
\eeq

The scaling laws~(\ref{vba}) and~(\ref{vfa}) generalize
(\ref{vblin}), (\ref{vflin}), (\ref{vbquad}), (\ref{vfquad}).
They can also be put in perspective with~(\ref{vhard}),
which holds in the situation where a hard bound $\ell$
is imposed on the distance $\abs{m}$ between both particles.
Let us introduce the length $\ell$
as the distance where the potential~(\ref{va}) equals unity, i.e.,
\beq
\ell=g^{-1/\a}.
\eeq
In terms of this length parameter,~(\ref{vba}) and~(\ref{vfa}) take the form
\beq
V^\B\approx2-\frac{C^\B}{\ell^{2\a/(\a+3)}},\qquad
V^\F\approx2-\frac{C^\F}{\ell^{2\a/(\a+3)}}.
\eeq
These expressions smoothly match their counterparts~(\ref{vhard})
in the $\a\to\infty$ limit, where the potential gets infinitely steep.
The exponent slowly converges to the limit value 2,
while the amplitudes~(\ref{cba}),~(\ref{cfa}) also converge to the limits~(\ref{chard}).

\begin{figure}[!ht]
\begin{center}
\includegraphics[angle=-90,width=.5\linewidth]{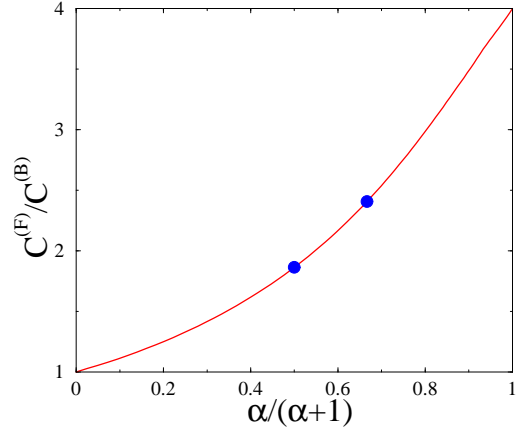}
\caption{\small
Plot of the universal amplitude ratio $C^\F/C^\B$ against $\a/(\a+1)$,
where $\a$ is the growth exponent of the confining potential.
Symbols: results~(\ref{cfb1}) and~(\ref{cfb2}) for $\a=1$ and $\a=2$.
Curve: data points obtained by means of a numerical calculation
of the eigenvalues $\E_0$ and $\E_1$ of~(\ref{scha}).}
\label{cbcf}
\end{center}
\end{figure}

The ratio $C^\F/C^\B$ can be viewed as a universal amplitude ratio.
This dimensionless quantity
depends only on the growth exponent $\a$ of the confining potential (see~(\ref{cfcba})),
increasing monotonically from 1 in the singular $\a\to0$ limit
to 4 in the $\a\to\infty$ (i.e.,~hard-bound) limit.
Its values for $\a=1$ and $\a=2$ have been given in~(\ref{cfb1}) and~(\ref{cfb2}).
Figure~\ref{cbcf} shows a plot of this ratio against $\a/(\a+1)$.

Finally, the amplitude ratio $C^\F/C^\B$ is always larger than unity.
More generally, bosonic bound states always have a larger maximal
spreading velocity than fermionic bound states in the same confining potential.
We shall come back to this very common property in section~\ref{discussion}.

\section{Discussion}
\label{discussion}

In this work we have investigated the continuous-time quantum walk performed
along a one-dimensional lattice
by bound states of two interacting particles.
The main focus has been on the profile
of the wavefunctions describing these bound states
in the center-of-mass coordinate,
and especially on the velocity characterizing their ballistic spreading
and on the structure of the whole profile,
which generically exhibits many internal fronts.

We have first revisited the problem of a single quantum walker
in a self-contained pedagogical fashion.
For the simple quantum walk where the particle hops to nearest neighbors only,
we have concentrated onto the dependence of the distribution
of the particle position on the initial state.
This distribution profile has generically two ballistic fronts.
Either one or even both fronts may be absent for carefully chosen special initial states,
as this was already the case for some examples of
discrete-time walks~\cite{sbk,tfm,iks,sbj}.
For the generalized quantum walk,
where the particle hops to first and second neighbors
with respective amplitudes 1 and $g$,
we have emphasized the possible occurrence of two internal fronts
in the distribution profile,
propagating at velocities $\pm V_-$, besides the two usual external ones,
propagating at velocities $\pm V_+$, which mark the endpoints of the allowed region
beyond which the wavefunction falls off exponentially.
The non-trivial dependence of the velocities $V_\pm$ on the amplitude~$g$
is in contrast with the equally spaced internal peaks
arising for the discrete-time quantum walks considered in earlier works~\cite{bca,mkk}.

The quantum walk of bound states of two bosonic or fermionic particles
has then been investigated in two situations:
either by imposing a hard bound $\ell$ on the distance between both particles,
or by generating the bound states by a smooth confining potential
growing as a power of the distance.
In both situations, we have focused
on the structure of the distribution of the center-of-mass coordinate.
We have investigated in detail the maximal velocities $V^\B$ and $V^\F$
characterizing the ballistic spreading of bosonic
and fermionic bound states, as well as the many internal fronts
of their distribution profiles.
In the case of a hard distance bound (section~\ref{shard}),
the maximal velocities have the simple expressions~(\ref{vbfres}).
The distribution profile generically exhibits
$\ell-1$ (resp.~$\ell-2$) internal fronts in the bosonic (resp.~fermionic)
case, besides the two extremal ballistic fronts.
In the situation of a smooth confining potential (section~\ref{ssoft}),
we have investigated in detail the cases of a linear and of a quadratic
(i.e.,~discrete harmonic) potential.
For all potentials of the form $W_m=g\abs{m}^\a$,
growing as a power of the distance between both particles,
the maximal velocities exhibit scaling laws of the form (\ref{vba}),~(\ref{vfa})
in the regime of a weak potential ($g\ll1$).
The associated amplitude ratio $C^\F/C^\B$ is universal,
in the sense that it depends only on the growth exponent $\a$ of the confining potential.
This ratio is larger than unity,
in accord with the more general property that
bosonic bound states have a larger maximal
spreading velocity than their fermionic counterparts in the same binding potential.

Let us now compare our findings with those of two recent papers~\cite{aam,qkg}
which are also devoted to the quantum walk of bound states.
The latter works deal with full two-body Hamiltonians,
along the lines of earlier investigations of the Anderson localization
of two interacting particles~\cite{ds,yi,kf}.
In such models bound states therefore coexist with a full two-particle continuum.
The situations studied in the present work
(hard distance bound or confining potential) only possess bound states,
so that the investigation of the latter is made easier.
As internal ballistic fronts are not discussed in~\cite{aam,qkg},
we shall henceforth concentrate on the maximal spreading velocity~$V$.
Reference~\cite{aam} deals with the antisymmetric (i.e., fermionic)
sector of a discrete-time model,
with a local interaction described by the action
of a special coin operator whenever both particles sit at the same site.
The associated coupling constant $g$ is an angle.
The analysis of the bound-state spectrum allows one to derive the maximal velocity $V$.
This quantity (whose expression is not explicitly given in~\cite{aam})
decreases continuously from $1/\sqrt{2}$ to $1/3$
as $g$ is increased from 0 to its maximal value of $\pi$.
Reference~\cite{qkg} describes the continuous-time dynamics of two identical particles
(either bosons or fermions) generated by a Hubbard-like Hamiltonian
with nearest-neighbor interactions.
The maximal velocities are derived by means of a perturbative approach
in the regime of very strongly attractive interactions.
Both spreading velocities are found to fall off to zero
as the inverse of the interaction strength,
and to obey the simple relation $V^\B=3V^\F$.

Even though very different models and regimes have been considered,
all the findings recalled above are consistent with the following universal characteristics
of the ballistic spreading of two-body bound states.
For a given interaction strength,
bosonic bound states always have a larger velocity of spreading
than their fermionic counterparts.
When the interaction strength is increased,
the spreading velocity decreases continuously from its free one-body value
down to zero or to a much smaller limiting value.

A natural extension of the present work consists in considering the quantum
walk performed by bound states of more than two identical particles.
A classical analogue consists of
multi-pedal molecular devices whose legs perform random walks,
known as molecular spiders~\cite{exp}.
Their behavior has been investigated theoretically,
both on a one-dimensional~\cite{akm} and a two-dimensional~\cite{ak} substrate.
The quantum version of the problem yields a special kind of $N$-fermion bound state,
which can be analyzed by techniques from integrable systems.
This will be the subject of future work~\cite{ustocome}.

\ack

It is a pleasure to thank R Balian, G Ithier and V Pasquier for fruitful discussions.

\appendix

\section{Stationary properties of a quantum system}
\label{app}

In this appendix we investigate the stationary properties of a finite quantum
system from a very general standpoint.
Consider a quantum system whose Hilbert space has finite dimension $N$,
working in a preferred basis $\ket{a}$ $(a=1,\dots,N)$.
In this basis, the Hamiltonian is given by an $N\times N$ Hermitian matrix $H$.
Assume that the energy eigenvalues $E_n$ $(n=1,\dots,N)$ are non-degenerate.
Let $\ket{n}$ be normalized eigenvectors, so that $H\ket{n}=E_n\ket{n}$.
Assuming the system is initially in state $\ket{a}$, we have
\beq
\ket{\psi(t)}=\sum_n\e^{-\ii E_nt}\ket{n}\braket{n}{a}.
\eeq
The probability of observing the system is state $\ket{b}$ at time $t$
reads therefore in full generality
\beq
P_{ab}(t)=\abs{\braket{b}{\psi(t)}}^2
=\sum_{m,n}\e^{\ii(E_n-E_m)t}\braket{b}{m}\braket{m}{a}\braket{a}{n}\braket{n}{b}.
\label{pab}
\eeq
We are interested in the stationary transition probabilities
\beq
Q_{ab}=\lim_{t\to\infty}\frac{1}{t}\int_0^tP_{ab}(t')\,\d t'.
\eeq
In the absence of spectral degeneracies, only diagonal terms $(m=n)$
in the double sum in~(\ref{pab}) contribute.
We thus obtain
\beq
Q_{ab}=\sum_n\abs{\braket{a}{n}}^2\abs{\braket{b}{n}}^2.
\eeq
The matrix $Q$ is real symmetric and positive definite.
It reads indeed\footnote{$R^T$ denotes the transpose of the matrix $R$.}
\beq
Q=RR^T,
\eeq
with
\beq
R_{an}=\abs{\braket{a}{n}}^2.
\eeq
The matrix $Q$ is non-trivial in general.
This is a manifestation of the well-known fact
that a finite isolated quantum system does not equilibrate,
in the sense that its stationary properties remember its initial state forever.

We now turn to the explicit example of a confined quantum walker,
i.e.,~a tight-binding particle
on a finite segment of $N$ sites labelled $a=1,\dots,N$.
The preferred basis is chosen to be local in space.
The Hamiltonian reads
\beq
\braopket{a}{H}{\psi}=\psi_{a+1}+\psi_{a-1},
\eeq
where $\braket{a}{\psi}=\psi_a$
and with Dirichlet boundary conditions $\psi_0=\psi_{N+1}=0$.
We have
\beq
E_n=2\cos\frac{n\pi}{N+1},\qquad
\braket{a}{n}=\sqrt\frac{2}{N+1}\sin\frac{an\pi}{N+1},
\eeq
($n=1,\dots,N$),
and so
\beq
Q_{ab}=\left(\frac{2}{N+1}\right)^2\sum_n\sin^2\frac{an\pi}{N+1}\sin^2\frac{bn\pi}{N+1}.
\eeq
This sum can be worked out explicitly.
We thus obtain
\beq
Q_{ab}=\frac{1}{N+1}\left(1+\frac12\delta_{a,b}+\frac12\delta_{a+b,N+1}\right).
\eeq

With respect to its uniform background value $1/(N+1)$,
the stationary probability~$Q_{ab}$ is thus enhanced by a factor 3/2
both at the starting point ($b=a$) and at the symmetric position ($b=N+1-a$).
In the particular situation where the starting point is the middle of an odd
segment ($N$ odd and $a=(N+1)/2$), the enhancement factor of the return probability
reaches 2.

The stationary mean value of the position $X$ of a walker launched at site $a$, i.e.,
\beq
\mean{X}=\sum_b bQ_{ab}=\frac{N+1}{2},
\eeq
is dictated by symmetry, and therefore independent of the initial state.
The corresponding variance,
\beqa
\mean{X^2}-\mean{X}^2&=&\sum_b b^2Q_{ab}-\biggl(\sum_b bQ_{ab}\biggr)^2
\nonumber\\
&=&\frac{N(N-1)}{12}+\frac{(N+1-2a)^2}{4(N+1)},
\eeqa
however shows a dependence on the initial position $a$ of the walker.

For $N=4$ we obtain
\beq
Q=\frac{1}{10}\pmatrix{3&2&2&3\cr 2&3&3&2\cr 2&3&3&2\cr 3&2&2&3}.
\eeq
The first row, i.e.,
\beq
Q_{11}=Q_{14}=\frac{3}{10},\qquad
Q_{12}=Q_{13}=\frac{1}{5},
\label{q4}
\eeq
agrees with the result~(\ref{pstat}) of a full dynamical analysis
of the two-fermion bound state with $N=\ell=4$.

\section*{References}

\bibliography{final.bib}

\end{document}